\documentclass[preprint,twocolumn,10pt,a4paper,byrevtex,prb,unsortedaddress,superscriptaddress,longbibliography]{revtex4-1}

\usepackage{amssymb}
\usepackage{natbib}
\usepackage{graphicx}
\usepackage{amsmath}
\usepackage{float}
\usepackage{verbatim}
\usepackage{color}
\usepackage{bbold} 
\usepackage{hyperref}
\begin{document}

\definecolor{Black}{rgb}{0,0,0}
\definecolor{Red}{rgb}{1,0,0}
\newcommand{\farkhad}[1]{\textcolor{Black}{#1}}
\definecolor{Blue}{rgb}{0,0,1}
\definecolor{Orange}{rgb}{1,0.55,0}

\newcommand{\diego}[1]{\textcolor{Black}{#1}}
\newcommand{\diegonew}[1]{\textcolor{Black}{#1}}

\title{Standing spin waves in Permalloy-NiO bilayers as a probe of the \diego{interfacial} exchange coupling}

\author{Diego Caso}
\affiliation{ 
Departamento de F\'isica de la Materia Condensada C03, Universidad Aut\'onoma de Madrid, Madrid 28049, Spain}
 
%

\author{Ana García-Prieto}%
\affiliation{Spanish CRG BM25-SpLine at the European Synchrotron, 71 Av. des Martyrs, 38000 Grenoble, France}
\affiliation{Instituto de Ciencia de Materiales de Madrid, Consejo Superior de Investigaciones Cient\'ficas, Madrid 28049, Spain}

\author{Eugenia Sebastiani-Tofano}%
\affiliation{Spanish CRG BM25-SpLine at the European Synchrotron, 71 Av. des Martyrs, 38000 Grenoble, France}
\affiliation{Instituto de Ciencia de Materiales de Madrid, Consejo Superior de Investigaciones Cient\'ficas, Madrid 28049, Spain}

\author{Akashdeep Kamra}%
\affiliation{Departamento de F\'isica Teorica de la Materia Condensada, C-05, Universidad Aut\'onoma de Madrid, Madrid 28049, Spain}
\affiliation{IFIMAC, Universidad Aut\'onoma de Madrid, Madrid 28049, Spain}

\author{Cayetano Hern\'andez}%
\affiliation{Departamento de F\'isica Aplicada, M-12, Universidad Aut\'onoma de Madrid, Madrid 28049, Spain}

\author{Pilar Prieto}%
\affiliation{Departamento de F\'isica Aplicada, M-12, Universidad Aut\'onoma de Madrid, Madrid 28049, Spain}
\affiliation{Instituto Nicol\'as Cabrera (INC), Universidad Aut\'onoma de Madrid, Madrid 28049, Spain}


\author{Farkhad G. Aliev}%
\email{farkhad.aliev@uam.es}
\affiliation{ 
Departamento de F\'isica de la Materia Condensada C03, Universidad Aut\'onoma de Madrid, Madrid 28049, Spain}
\affiliation{IFIMAC, Universidad Aut\'onoma de Madrid, Madrid 28049, Spain}
\affiliation{Instituto Nicol\'as Cabrera (INC), Universidad Aut\'onoma de Madrid, Madrid 28049, Spain}

\date{\today}
\begin{abstract}
Ferromagnetic/Antiferromagnetic (FM/AFM) bilayers dynamics have been a recent topic of interest due to the interaction occurring at the interface, where the magnetic moments of the AFM can be imprinted into the FM, and the exchange bias field can affect these dynamics. Here, we investigate Permalloy (Py) and NiO (Py/NiO) hybrids and for comparison single Py films in the broad Py thickness range varied from few nm to 200 nm by using static (Kerr effect) and dynamic (spin waves) measurements along with micromagnetic simulations. We observe hybrid modes between uniform (ferromagnetic resonance FMR, n=0) and perpendicular standing spin waves (PSSWs, n=1, 2) and a clear enhancement of the PSSWs modes frequencies upon interfacing Py with NiO both from experiments and simulations. This enhancement becomes less pronounced as the thickness of the film increases, demonstrating its interfacial origin rooted in the exchange coupling between the FM and AFM layers. Besides, through micromagnetic simulations, we investigate and correlate changes in spatial profiles of the PSSWs with the interfacial exchange coupling. As the thickness is increased, we see that the n=1 and n=2 modes begin to couple with the fundamental FMR mode, resulting in asymmetric (with respect the Py layer center) modes. Our results suggest that PSSWs detection in a ferromagnet offers a means of probing the interfacial exchange coupling with the adjacent AFM layer. Furthermore, the controlled spatial symmetry breaking by the AFM layer enables engineering of PSSWs with different spatial profiles in the FM.

\end{abstract}

\maketitle

\section{\label{sec:level1}Introduction}

Over the last few decades, one focus in the investigation of ferromagnet-antiferromagnet bilayers has been concentrated around the induction of the uniaxial anisotropy asymmetry due to  the influence of exchange bias \cite{Nogues1999, Smardz2002, Blachowicz2021}
Recently also an opposite limit has been explored when a thin ferromagnet is coupled to a much thicker epitaxial antiferromagnet \cite{Bommanaboyena2021} to operate as \farkhad{a readout mechanism} for antiferromagnetic spintronics.

An important factor present in both limits is the exchange coupling at the AFM/FM interface, since it not only shifts the static magnetization hysteresis curves \cite{Smardz2002}, but may also affect magnetization dynamics. For example, exchange coupling to antiferromagnets was used to increase the zero field FMR frequency \cite{Phuoc2010}. On the other hand, excitation of spin waves (SWs) in a ferromagnet coupled to an antiferromagnet was proposed as a method to propagate spin excitation into the antiferromagnet \cite{Jenkins2020}. Exchange field was reported to influence the lowest frequency uniform FMR mode \cite{Stamps1996} but could even have an stronger influence in the short wavelength perpendicular standing spin waves (PSSWs) when magnetization symmetry in the direction perpendicular to the interfaces is broken.

Recently, PSSWs have been studied in other symmetry breaking systems such as heavy metal/magnetic insulator (HM/MI) bilayers, where the frequency of the standing wave modes experiences a shift \cite{Lee2023}. The effect was attributed to an additional confinement of the waves upon a local increase of the interfacial damping due to spin pumping \cite{Lee2023}. 
A number of other recent experiments also used PSSWs to probe coupling in ferromagnetic bilayers \cite{Kostylev2009, Qin2018, Zhang2021, Kennewell2010} or individual surface anisotropies on epitaxial ferromagnetic layers with nonequivalent interfaces \cite{Solano2022}. Interestingly, the PSSW modes excited in two interacting ferromagnetic films have been proposed to enable quantum entanglement between distant magnon sources for quantum information applications \cite{Sheng2023}.  

Application of PSSWs to explore and understand interfacial spin dynamics in AFM/FM bilayers has been however much less explored. To our best knowledge, such studies have only been done by varying AFM thickness (fixing thickness of FM layer) and only by using a single microwave excitation frequency \cite{Magaraggia2010}. As regards broadband magnetization dynamics studies, they have been carried out only for the strongly coupled AFM/FM bilayers, where the FMR mode is influenced and split into two separated modes by the adjacent AFM \cite{Hamdo2023}. Interestingly, recent studies also discovered the appearance of an additional (to the uniform FMR and PSSWs modes) low frequency (below GHz) modes in the coupled AFM/FM bilayers, attributed to the excitation of the interfacial layers with reduced magnetic moment \cite{Polishchuk2021}. The emergence of an additional mode at the interface has also been discussed in a ferrimagnet/ferromagnet bilayer \cite{Luthi2023}. 
\farkhad{These recent advances demonstrate a high potential for investigating and engineering spin dynamics in AFM/FM bilayers, which is the goal of the present article.}

As to the 
\farkhad{possible interaction between two different magnon modes},
it was Grünberg et al. \cite{Grunberg1982} who first predicted that Damon Eshbach (DE) surface modes would not interact with PSSWs in the dipolar limit. However, if large exchange was considered, the modes do interact repulsing each other and causing the elimination of the crosspoints \cite{Grunberg1982}. On the other side, Klingler et al. reported anti-crossings of YIG PSSWs and the Co uniform precession FMR mode \cite{Klingler2018}, modeled by a mutual spin pumping and exchange interaction \cite{Gou2023, KumarMondal2023}. Generally, the investigation of magnon-magnon coupling in the same material and with a shared propagation direction such as PSSW of different order remains under-explored. To our best knowledge, only Wang et al. reported modes anti-crossings for interactions in quasi-uniform \farkhad{in-plane propagating dipole-exchange spin waves} modes with first order PSSWs in a FM/AFM heterostructure \cite{Wang2022}.

\begin{figure}[t]
  \centering
  \includegraphics[width=1\linewidth]{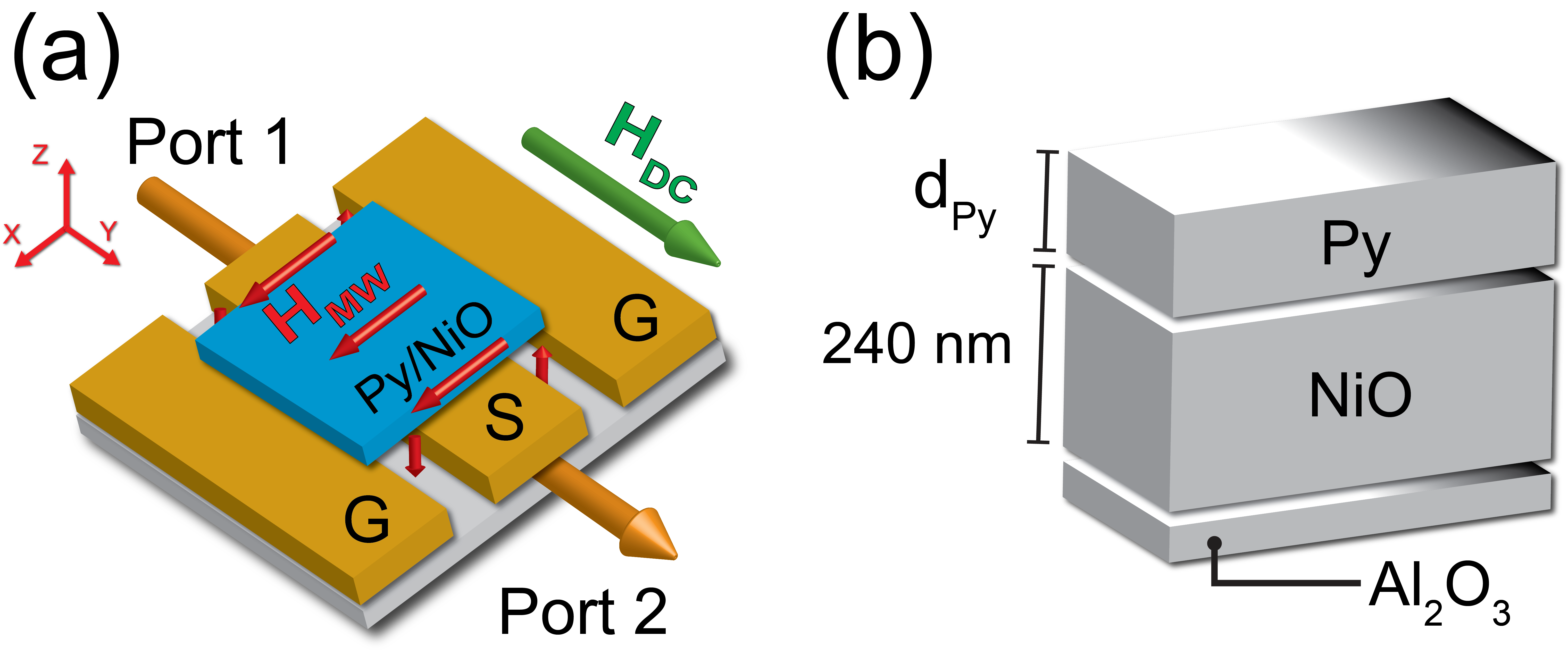} 
  \caption{(a) Sketch of the experimental setup, depicting the MW and DC field distributions on the film placed on the coplanar waveguide (CPW). The high frequency signal travels from Port 1 to Port 2 through the waveguide (S), in between the two grounds (G). (b) Sketch of a typical Py/NiO film investigated here.}
  \label{setup}
  \end{figure}
  
\begin{figure*}[t]
  \centering
  \includegraphics[width=0.65\linewidth]{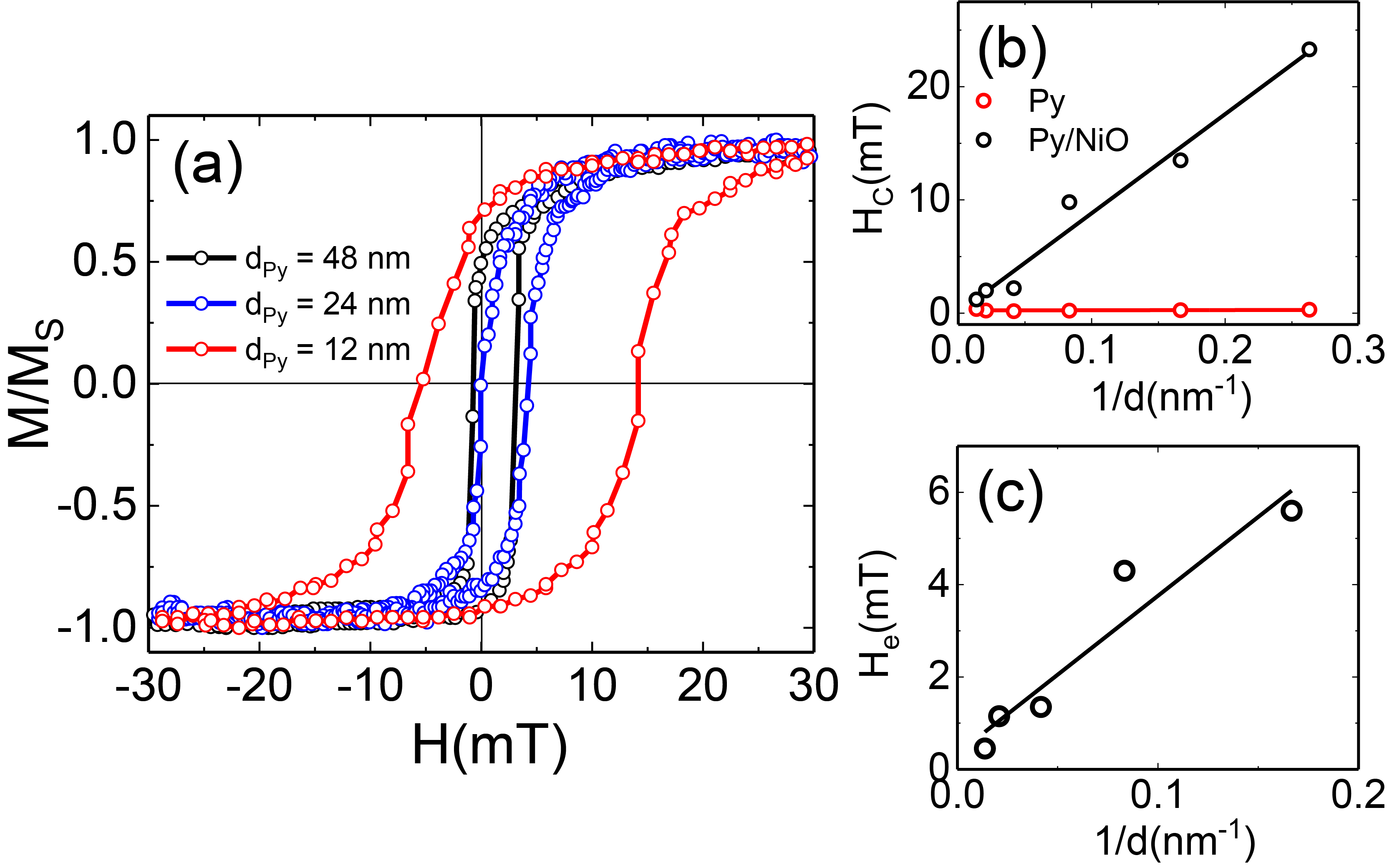} 
  \caption{(a) Hysteresis loops for some of the used Py/NiO films on its easy axis, the Py thickness is depicted in the legend. (b) Coercive field of all of our films. (c) Exchange bias of the films extracted from the asymmetry of the hysteresis loops.}
  \label{hysteresis}
  \end{figure*}
  
Here, we investigate the influence of the exchange field on the static and dynamic response in Py(d\text{$_{Py}$})/NiO(\diego{240} nm) bilayers in the broad frequency (up to 30 GHz) and Permalloy thickness (24-200 nm) ranges (\diego{details of the sample growth and characterization are given in Appendix \ref{appendix: characterization}}). \diego{The thickness of the NiO layer was high enough to
avoid AFM thickness effect}. We observe both the emergence of the surface layer's magnetization dynamic response and the influence of the exchange coupling on the perpendicular standing spin waves in Permalloy. \farkhad{Our observation of frequency shifts in the PSSW modes resulting from the interfacial exchange offers a dynamic means to probe and characterize the interface properties. Furthermore, via micromagnetic simulations, we demonstrate the ability to engineer the PSSW modes' spatial profiles via the interfacial exchange.}
\section{Methods}
\label{methods}

\subsection{High frequency experimental setup}
\label{ex_setup}
To probe the dynamic magnetic properties of the different investigated films, we use a basic vector network analyzer (VNA) high frequency setup (see for details Ref. \cite{Aliev2011}). By placing the films directly on top of a 50$\Omega$ coplanar waveguide (CPW) connected to the VNA, the related small MW field will oscillate in the same plane as the sample (see Figure \ref{setup}(a)). A DC magnetic field up to 2.7 kOe is simultaneously applied perpendicularly to the microwave field to generate torque in the magnetic moments of the film. The measurements were made by sweeping the frequency for constant DC magnetic fields. Microwave frequency was swept from 10 MHz up to 30 GHz and MW power was set at 0 dBm. The scattering matrix element S$_{21}$, which relates the magnitude and phase of the input signal sent to the CPW and the received signal, is an indicative of the microwave permeability in the system \cite{Muller2021}. The real part of S$_{21}$ is then calculated and normalized at maximum DC field to obtain the frequency-applied field maps. 

\subsection{Simulation details}
\label{sim_details}
Dynamic simulations were carried out using the Mumax\text{$^3$} \cite{Mumax2014} software for micromagnetic simulations. The discretization cell size was set at 2 nm $\times$ 2 nm $\times$ 1 nm, which is far below the Permalloy (Py) exchange length (5.3 nm) \cite{Wang2017}. The decreased cell size in the $z$ direction is deliberately imposed to obtain an enhanced precision on this coordinate and accurately characterize the thickness modes. \diego{The size of the simulated structure was \text{128 nm$\times$ 128 nm $\times$ \text{d$_{Py}$}}, and periodic boundary conditions (PBCs) were imposed on the in-plane dimensions to avoid edge effects.} We used typical magnetic parameters of Py in our simulations \cite{Lara2013}:
$M_{s} = 1.16\, T;\quad A_{ex} = 14\times 10^{-12}\, J/m; \quad \alpha = 0.001 \quad,$ where M$_s$ is the saturation magnetization of the Py and A$_{ex}$ and $\alpha$ specify the exchange stiffness constant and the damping constant respectively. 

Lower \text{$M_{s}$} is set near the top surface of our films \diego{(0.25 T on the 2 top layers)}. This translates to a lower demagnetizing field and a higher effective potential in these layers, thus enabling the pinning of the PSSWs modes \cite{Waring2023}. This effect is plausible due to the surface oxidation typically experimented by ferromagnetic films \diego{or a magnetization symmetry breaking in the normal direction due to a capping layer}. For Py specifically, the thickness of this surface oxidation layer has been reported to be close to 1.5 nm \cite{Fitzsimmons2006}. In the case of the Py films, the variation of \text{$M_{s}$} is the most conceivable way to induce changes in the surface pinning parameter \cite{Waring2023}, therefore generating magnetization symmetry breaking and allowing the excitation of PSSWs. However in the FM/AFM system, the simulations include an interfacial surface anisotropy (\text{$K_s$}) due to the influence of the exchange bias. In that case, the magnetization symmetry breaking is accomplished on both surfaces of the films.

In order to uncover the main SW modes, a homogeneous magnetic field pulse of the following form: $\vec{H}_{pulse} = H_0 sinc\left(\frac{2\pi t}{t_0}\right)\hat{u}_x$ was employed over a 10 ns duration. Here, H$_0$ represents the amplitude of the applied field pulse, 20 Oe, and t$_0$ denotes the half width of the pulse, 10$^{-12}$ s. The used excitation is homogeneous over the thickness of the material thickness. The averaged out-of-plane (OOP) magnetization of the system was converted into the frequency domain after the use of a Fourier Transform \cite{Park2003}. The process was repeated for each of the presented magnetic fields. 

\section{Experimental results and discussion}
\subsection{Static characterization of the Py and Py/NiO films}
  
\begin{figure*}[t]
  \centering
  \includegraphics[width=0.75\linewidth]{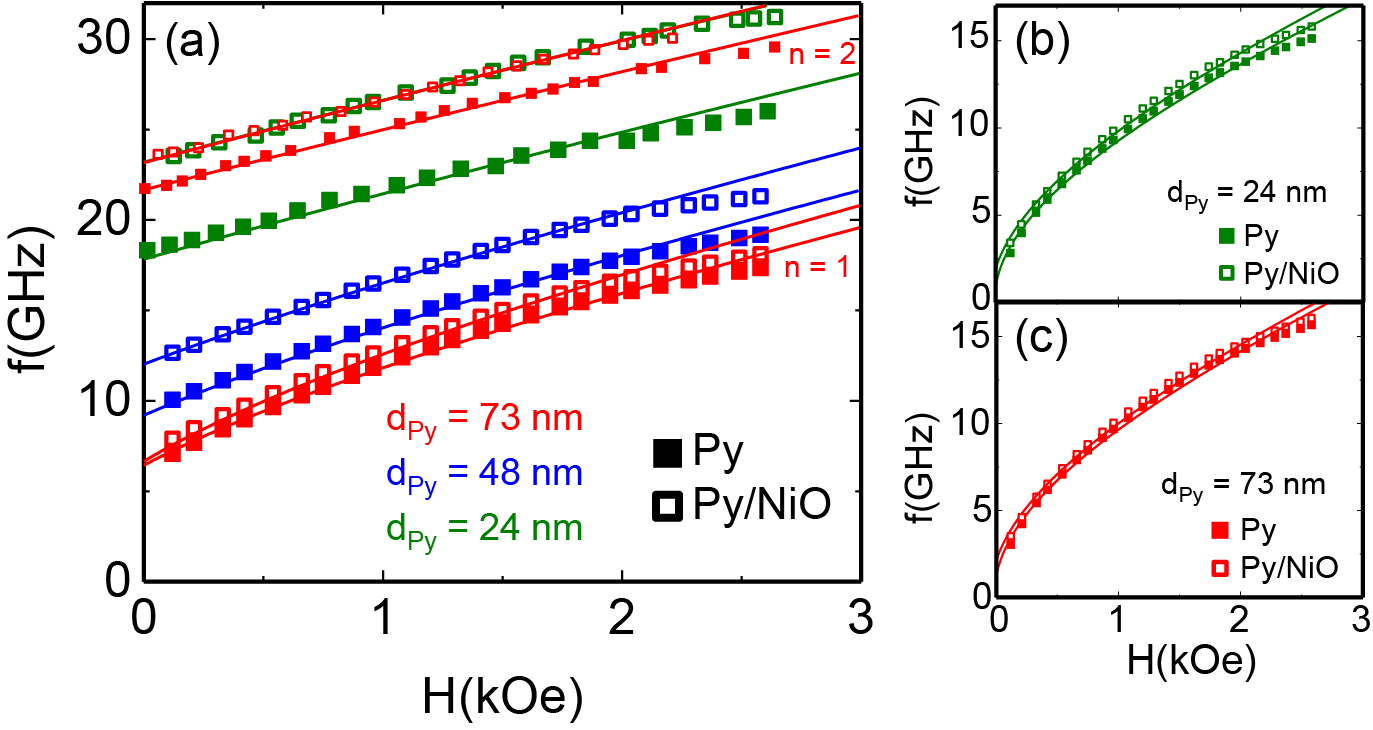} 
  \caption{(a) PSSWs on the Py and Py/NiO for 73, 48 and 24 nm thick Py against frequency and bias field. 
  (b) and (c) represent the n = 0 mode of the 73 and 24 nm films respectively against the magnetic field. Lines in (a, b, c) are the corresponding fits to Eq. \ref{eq1}.}
  \label{PSSW}
  \end{figure*}
  
Kerr magneto-optical experiments (MOKE, see Appendix \ref{appendix: characterization}) reveal a clear enhancement of the coercive field of all of our ferromagnetic films upon the addition of NiO (see Figure \ref{hysteresis}(b)). This phenomenon has been documented in both experimental and analytical studies of FM/AFM films \cite{Mauri1987, Li2000, Hu2003}. Moreover, as the experimental data suggests, this effect is suppressed once the thickness of the FM layer becomes large enough, which can be attributed to the reduced effect of the interface exchange coupling strength relative to the overall magnetization of the film.

The shift in the hysteresis loops from a field-centered position indicates the presence of an exchange bias field in the Py/NiO films. The interplay between the exchange bias and the thickness of the film is depicted in Fig. \ref{hysteresis}(c), yielding also in a reduction of the field as the Py layer becomes thicker, once again, an effect related to the reduced influence on the overall film magnetization. These findings collectively point to the existence of a quantifiable exchange coupling within our FM/AFM films. 

\subsection{Effect of the NiO on the frequency of the observed standing waves}
\label{subsec:NiO_freq}

Performed VNA-FMR experiments (see Methods section \ref{ex_setup}) in both sets of Py and Py/NiO films revealed the presence of an uniform FMR mode, characteristic of the FM layers. For films of 24 nm and above, we also found other set of higher order spin wave resonances above the frequency of the FMR mode. These spin wave modes are identified as exchange-dominated thickness-dependant modes related to perpendicular standing spin waves (PSSWs) in our films \cite{Dbrowski2021}. Several factors contribute to the identification of the detected modes as PSSWs. For instance, the dependence of the modes with the thickness of the films is a clear indicator, as well as the alignment with the performed simulations and the extensive previous literature and reports of these modes on thick ferromagnets.


For all of our films, the presence of the NiO layer shifted the standing spin wave modes to higher frequencies (see Figure \ref{PSSW}(a)). Our experimental data indicates that this frequency shift decreases with the enhancement of the thickness of the Py layer, as depicted in Figure \ref{Shift}. This suggests a direct relation between the magnitude of the exchange coupling effects on the overall behavior of the magnetic moments in the film as the FM layer grows thicker and the frequency shift of the PSSW modes experienced when the NiO layer is added. We notice that the FMR mode, which is not thickness-dependent, also experiences a slight frequency shift in the same direction for all of our pair of samples when the NiO is added, as Figures \ref{PSSW}(b, c) illustrate.

As briefly mentioned in the introduction, we believe that the frequency shift experienced by our films is a direct consequence of the Py magnetic moments pinning at the interface due to the exchange bias. This absence of free spins results in an additional confinement of the PSSWs. Furthermore, an estimation of the thickness reduction of the Py layer due to the pinned magnetic moments is presented in Figure \ref{Shift}(b), by letting the thickness variable to be free in Equation \ref{eq1}. A 2 nm consistent effective thickness reduction can be appreciated for all of the films. This additional confinement is further supported by the experiments done on extremely thin films (see Appendix \ref{appendix:low_thickness}). 

The resonances of the PSSW modes were initially predicted by Kittel \cite{Kittel1958}, and Seavey et al. \cite{Seavey1958} were the first to experimentally observe PSSW modes. The following dispersion relation \cite{Gui2007, Conca2014, Navabi2017, Qin2018, Liu2020} has been broadly used to characterize the standing spin waves of a ferromagnetic material assuming perfectly rigid boundary conditions as an extension of the Kittel formula \cite{Kittel1948}:
\begin{align} \label{eq1}
\begin{split}
 f_{P S S W}=\frac{\gamma}{2 \pi} \sqrt{\left[(H_{e x t}+H_k)+
 \frac{2 A_{e x}}{M_s}\left(\frac{n \pi}{d}\right)^2 \right]\ldots} \\ 
 \overline{\ldots\times\left[(H_{e x t} + H_k) +\frac{2 A_{e x}}{ M_s}\left(\frac{n \pi}{d}\right)^2+4\pi M_s\right]}
\end{split}
\end{align}
Where n is the quantized order of the PSSW mode along the thickness direction (n=0 for the FMR mode), $d$ the thickness of the film, \text{A$_{ex}$} the exchange stiffness constant, \text{M$_s$} is the saturation magnetization, which is obtained by fitting the uniform mode, \text{H$_{ext}$} the external magnetic field, \text{$\gamma$} the gyromagnetic ratio and \text{H$_k$} corresponds to the in-plane uniaxial anisotropy field.

\begin{figure*}[t]
  \centering
  \includegraphics[width=0.8\linewidth]{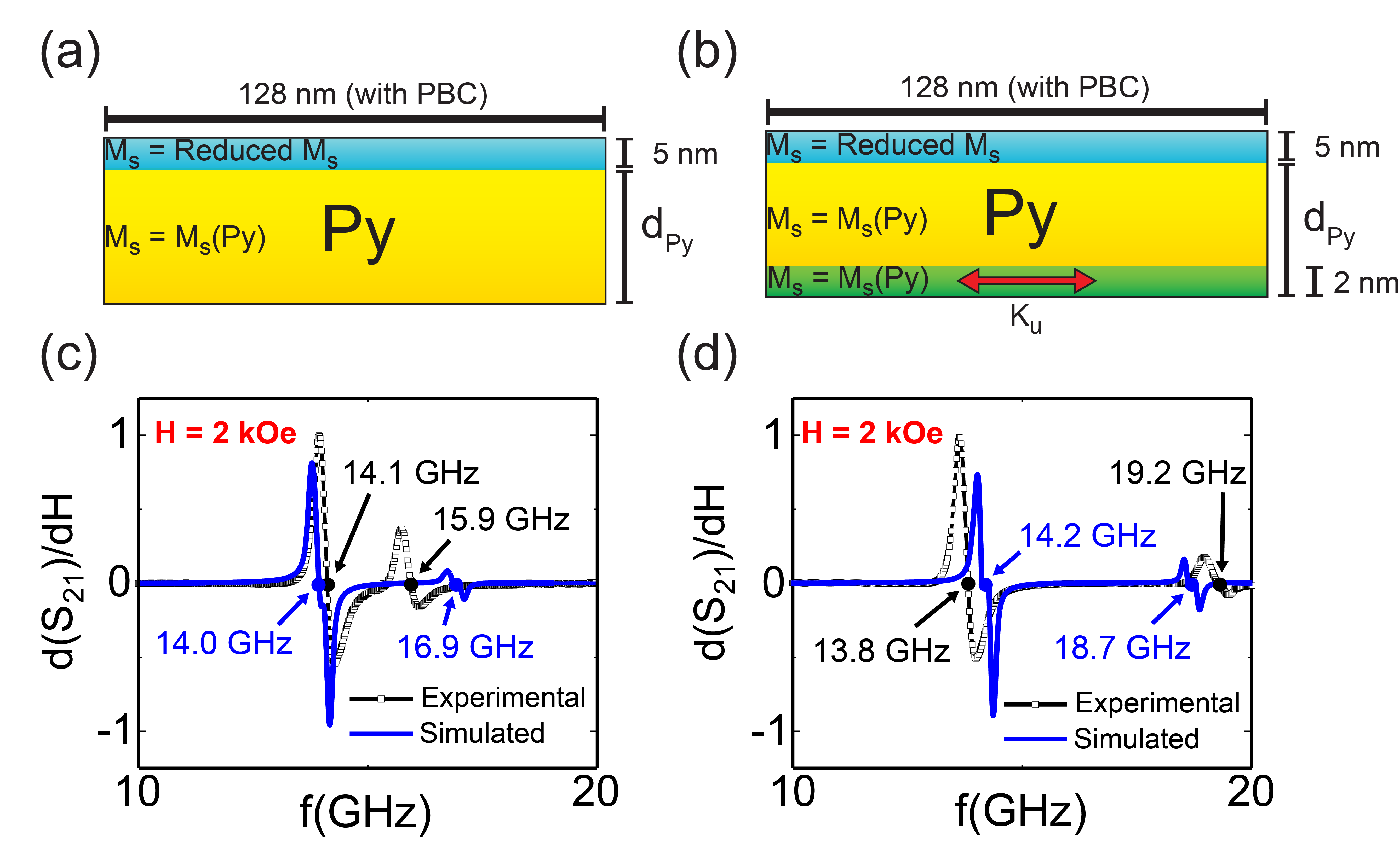} 
  \caption{(a) Sketch of a cross-section in the case of the simulated Py film, and (b) introducing a FM/AFM interface. Simulated and experimental results via VNA-FMR of the field derivative of the transmission at 2 kOe in the case of Py(48 nm), (c), and Py(48 nm)/NiO, (d) with an IP uniaxial anisotropy of K$_u$ = 10$^{-5}$ J/m$^3$. For the simulated Py(48 nm)/NiO film, an exchange bias field of 1 kOe IP was manually introduced in the bottom layer of the system.}
  \label{simulations}
  \end{figure*}
  
\begin{figure}[t]
  \centering
  \includegraphics[width=1\linewidth]{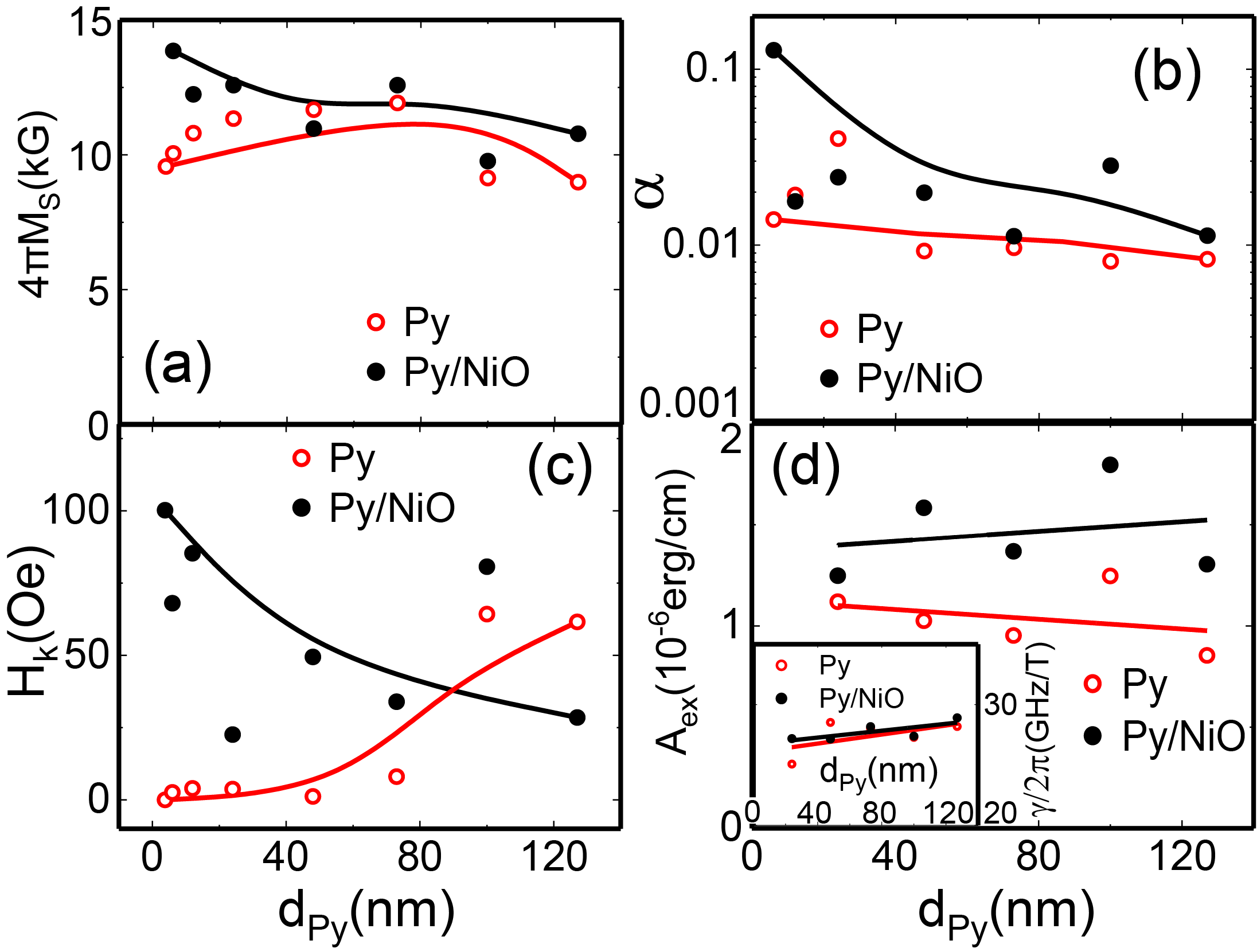} 
  \caption{Magnetic parameters of the Py/NiO bilayers and control Py samples. (a) Saturation magnetization of the films. (b) Gilbert damping. (c) Calculated IP anisotropy field. (d) Exchange stiffness constant calculated following a fit to Eq. \ref{eq1}. Inset in (d) corresponds to the gyromagnetic ratio, also obtained as a free parameter from a fit to Eq. \ref{eq1}.}
  \label{magnetic_parameters_fit}
  \end{figure}

The experimentally detected PSSWs modes in our films were fitted to eq. \ref{eq1}. Solid lines in Figure \ref{PSSW}(a, b, c) prove a great accuracy of the fit. Figure \ref{magnetic_parameters_fit}(a, b, c, d) represent the magnetic parameters obtained by fitting the PSSWs. While the saturation magnetization remains very consistent at all thicknesses, the anisotropy field of the Py/NiO samples progresses similarly as the exchange bias and coercive fields extracted from the static hysteresis loops. This is expected since the exchange bias of the films have less influence on the overall magnetization as the films grow thicker (Figure \ref{hysteresis}(c)). In the control samples of Py, however, the anisotropy field remains practically constant for low thicknesses, proving once again the quantifiable aspect of the interface coupling between the Py and the NiO in our study. 

The damping parameter was obtained by fitting the FMR linewidth (\text{$\Delta H$}) to the following formula:
\begin{equation}
    \Delta H = \Delta H_0 + \alpha \frac{4\pi}{\gamma}f
\end{equation}

Where \text{$\alpha$} is the Gilbert damping, \text{$f$} corresponds to the resonance frequency, and \text{$\Delta H_0$} is the intrinsic damping contribution, which denotes the frequency independent linewidth coming from the film inhomogeneities. \diegonew{The Py/NiO samples exhibit a higher magnetic damping (almost an order of magnitude for smaller Py thickness) in comparison with the control Py samples, as depicted in Figure \ref{magnetic_parameters_fit}(c). The damping parameter could be influenced by the spin pumping to antiferromagnet \cite{Wang2020} and exchange bias at the interface, freezing the magnetic moments of the Py and preventing larger precessions.}

\begin{figure*}[t]
  \centering
  \includegraphics[width=0.95\linewidth]{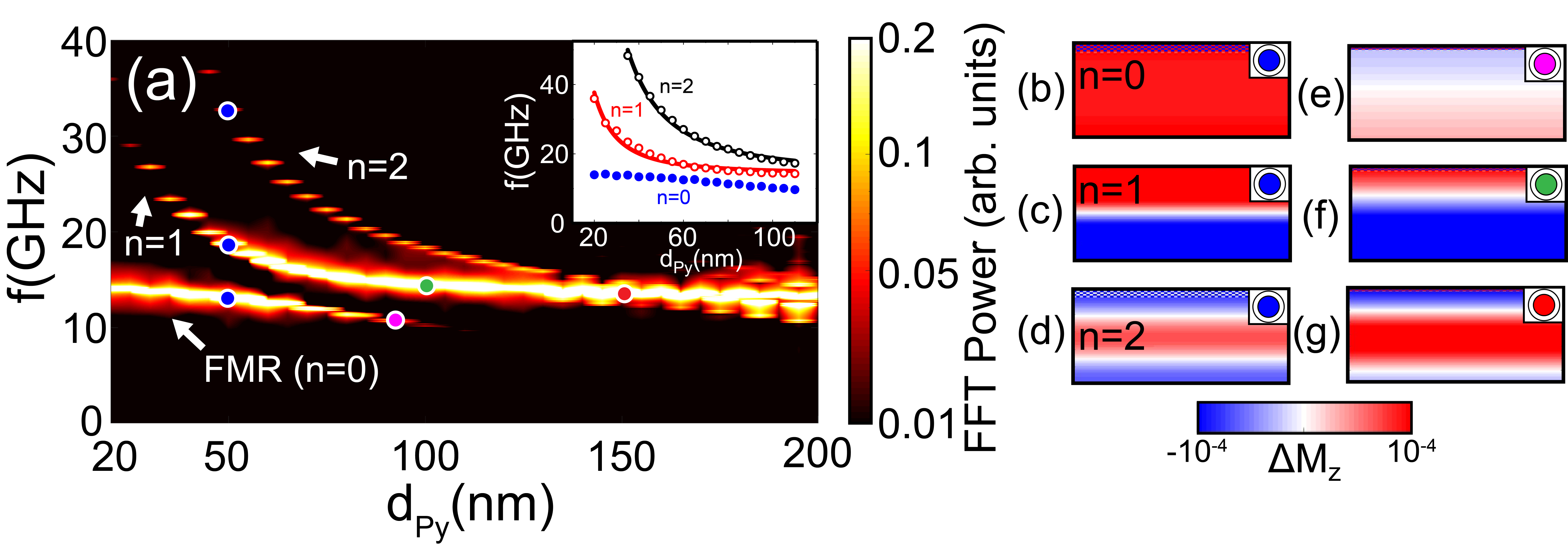} 
  \caption{(a) Evolution of the primary SW mode frequencies at a constant magnetic field of H = 2 kOe as the Py film thickness varies from 20 nm to 200 nm. The FFT power is normalized at each film thickness. The inset exhibits the three detected modes until 110 nm, where the lines in the n=1 and n=2 curves represent a fit of the behavior of the modes using the equation f(GHz) = A + B/d$_{Py}^2$(nm), A and B being free parameters. (b, c, d) Snapshots of the n=0, n=1 and n=2 PSSWs modes respectively at 50 nm of thickness. (e) Snapshot of the hybrid n=0 mode at 90 nm (f). Snapshot of the asymmetric main mode at 100 nm. (g) Snapshot of the main mode at 150 nm.}
  \label{thickness}
  \end{figure*}
  
\section{Numerical Results of PSSW presence on Py films}
\label{sec:numerical results}

\farkhad{In order to deeper understand the magnetization dynamics} we numerically investigate via micromagnetic simulations (see the Methods section \ref{sim_details} for details) the presence of PSSWs on 20-200 nm-thick Py films. The films are 
thick enough to accurately resolve the PSSW modes (the number of cells in the OOP direction is directly correlated to the thickness of our simulated samples) and to observe their possible shifts in the frequency when the interaction with the AFM is introduced. 

With a surface magnetization gradient (see Methods section \ref{sim_details}) caused by a possible oxidation and/or capping, we aim to manipulate the interface pinning parameter, which strongly affects the spin configuration of the PSSW modes across the film \cite{Waring2023}. \diego{We disregard surface anisotropies in our simulations}. The pinning parameter has the following form:
\begin{equation}
    \xi = \frac{2K_s}{qM_s^2}
    \label{pinning}
\end{equation}
where $K_s$ is the surface anisotropy, $M_s$ the saturation magnetization, and $q =\nabla^2 \textbf{M}$ is related to the non-uniform part of the exchange field. Most of the studies involving PSSWs use the case of surface rigid pinning ($\xi \rightarrow \infty$) introduced by Kittel \cite{Kittel1958}, in which equation \ref{eq1} is a solution. The other limit case, no surface pinning ($\xi = 0$), in which the magnetic moments are free in the surface is the other instance under extensive study \cite{Ament1955}. However, experimentally, in both of these limits it would be difficult to excite the n = 1 PSSW since there is no net magnetization in the mode profiles \cite{Gurevich2020}.

Waring et al. \cite{Waring2023} recently demonstrated that dynamic pinning, i.e., $\xi$ being a nonzero value, describes better the behavior of the excited PSSWs without the need to see major changes in the exchange stiffness constant \text{$A_{ex}$} from the fundamental mode, n=0, to higher order PSSWs (n=1, 2...). On this topic, some works have introduced magnetization graded ferromagnetic films to generate perpendicular standing spin wave modes in their simulations \cite{Gallardo2019, Waring2023}. This practice accomplishes an enhancement of the pinning on the surface in which the \text{$M_s$} gradient is introduced, see Eq. \ref{pinning}.

Our micromagnetic simulations performed on a Py layer with an enhanced pinning created by the reduction of the saturation magnetization on the top layers revealed field-dependent modes (see Figure \ref{simulations}(b)). These modes resemble the frequency of the PSSWs observed experimentally, as the blue and black lines suggest for the 48 nm Py sample in Figure \ref{simulations}. In fact, a full spectrum over the thickness of the films from 20 to 200 nm (Figure \ref{thickness}) reveals n = 0, 1 and 2 (and weak n = 3, 4, 5, see Appendix \ref{appendix:nabove2}) PSSW modes, which appear at every thickness, 20 nm, and 55 nm respectively for the frequency range that is explored in our VNA-FMR experiments (up to 30 GHz). These results match with the experimentally observed modes in our Py films, see Figure \ref{PSSW}(a), which show n = 2 modes only in the case of 73 nm thick Py and the emergence of the first order PSSW only in the films of 24 nm and above. Weaker modes above 30 GHz are not observed experimentally in our system due to the strong losses experienced at high frequencies.

The detected n$>$0 modes in our simulations are proportional to 1/d$^2_{Py}$, as seen in the inset of  Figure \ref{thickness}. This behavior is indicative of PSSWs described by Eq. \ref{eq1} and agrees with the reports on PSSWs in other FM films \cite{Grunberg1982}. On the other hand, the frequency of bulk (FMR) mode remains almost constant when the thickness of the film is increased, slightly decreasing in frequency once the frequency of n=1 mode approaches the fundamental mode. Eventually, the two modes combine at d$_{Py}>$110 nm.  

\subsection{Simulating the Py/NiO interface. Optimization of the uniaxial anisotropy}
\label{optimization}
The simulations of magnetization dynamics in FM/AFM bilayer were performed by introducing an extra layer in the system as an uncompensated AFM (see Figure \ref{simulations}(b)). This produces uncompensated spins at the interface that lead to an exchange bias between the FM and the AFM \cite{DeClercq2016, DeClercq2017}. Such approach is valid as spin dynamics observed experimentally is related to the FM layer interacting with an uncompensated antiferromagnet (NiO). 

\begin{figure}[t]
  \centering
  \includegraphics[width=0.75\linewidth]{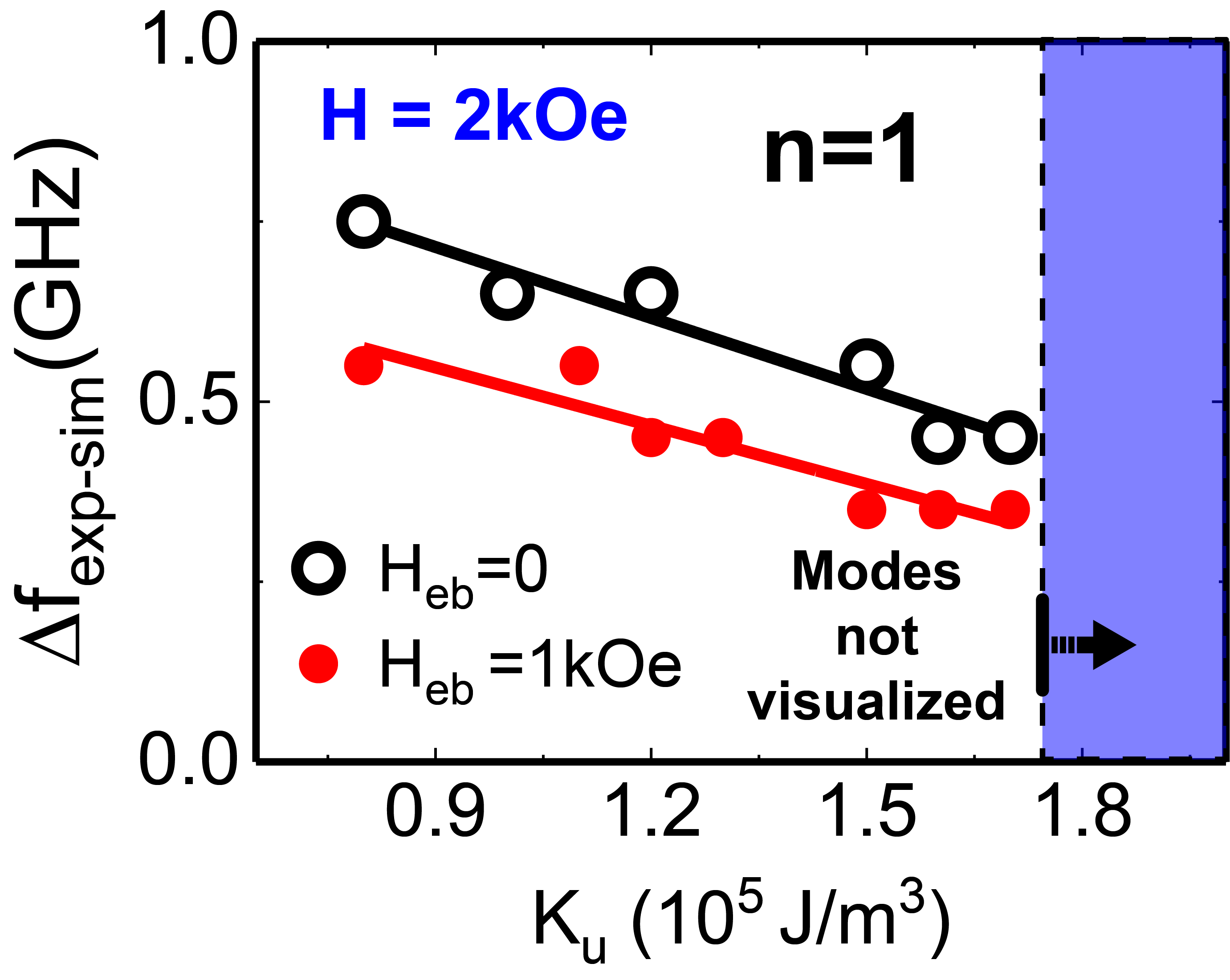} 
  \caption{Difference in frequency of the first PSSW mode (n=1) for the Py(48 nm)/NiO film obtained experimentally via the dynamic VNA-FMR experiments and the simulated Py(48 nm) film with an antiferromagnetic interface in the bottom. The darker region in the figure indicates the limit where the PSSWs modes are not excited due to the high K$_u$ affecting the ground state configurations.}
  \label{anisotropy}
  \end{figure}

Shifts in the frequency of the modes, especially in the experimentally observed first order PSSW (n=1), are also present in all of the performed simulations featuring an AFM/FM interface. These alterations move towards higher frequencies upon introducing the AFM, which also aligns with the empirical results. Specifically, the uniaxial anisotropy introduced in the interface has a direct influence on the frequency of the PSSWs. As depicted in Figure \ref{anisotropy}, a greater uniaxial unisotropy induced by the AFM leads to higher frequency modes. This trend continues until the system reaches a threshold (K$_u>$1.7$\times$10$^5$ J/m$^3$) where the system relaxes into different ground state configurations, rendering the detection of PSSWs no longer feasible. In our simulations employing solely uniaxial anisotropy, for a 48 nm thick film, a difference in the first order perpendicular standing wave of less than 0.5 GHz for uniaxial anisotropy constants of 1.5-1.7$\times$10$^5$ J/m$^3$ was found between the experimental and the simulated results. However, the frequency of the modes can be further enhanced by introducing a local in-plane exchange bias field in the bottom layer of the order of H$_{eb}$= 1 kOe. This enhancement brought the modes closer to the experimentally observed values by less than 0.4 GHz, as illustrated in Figure \ref{anisotropy}. 
The significant discrepancy between the exchange bias values obtained in simulations of the FM/AFM heterostructure, notably an order of magnitude greater than the exchange bias detected via MOKE (see Figure \ref{hysteresis}(c)), is rationalized by recognizing that MOKE spectroscopy specifically probes the top surface of the sample. However, the exchange bias effect is interfacial, as accounted for in the simulations, making it effectively larger than the measurements taken several nanometers above (see Appendix \ref{appendix:exch bias}).

\begin{figure}[t]
  \centering
  \includegraphics[width=1\linewidth]{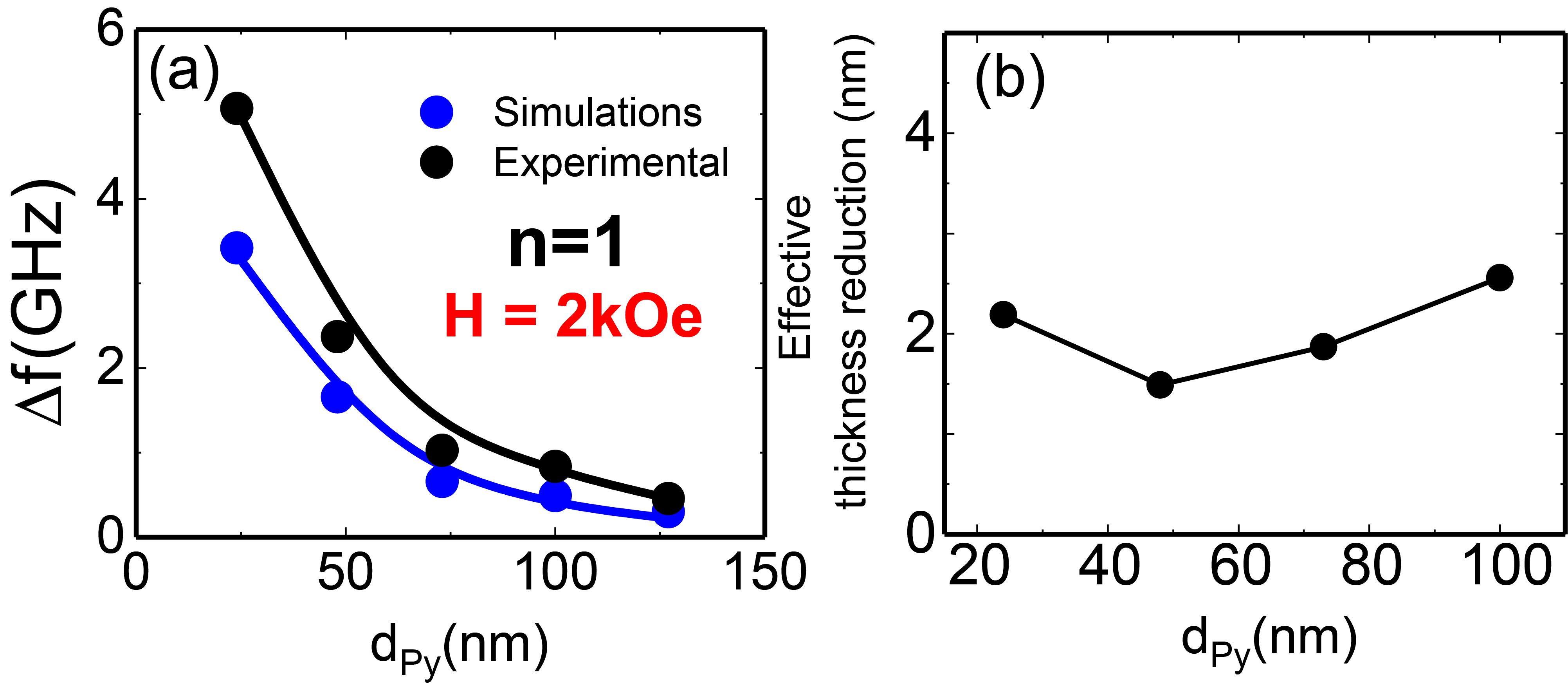} 
  \caption{(a) Frequency shift of the first order PSSW \diego{against the thickness of our} films upon the addition of the NiO layer at an applied magnetic field of 2 kOe and simulated frequency shift at the same thicknesses and field for a Py film with an antiferromagnetic interface. \diego{(b) Effective thickness reduction of the FM film upon the addition of NiO against the thickness of the Py layer.}}
  \label{Shift}
  \end{figure}
Additionally, the frequency shifts of the first order PSSWs are tracked at a constant magnetic field of 2 kOe for each of the films (see Figure \ref{Shift}). As the thickness of the films decreases, we notice a decline in the frequency shifts due to the decreasing impact of the AFM/FM interface on the overall magnetization of the film in our experiments, which is to be expected. Our simulations also yield similar results, showing a consistent trend of diminishing the shifts upon increasing the thickness of the Py film, thereby confirming a valid interpretation of the Py/NiO system. \diego{To examine the evolution to the main SW modes in the Py/AFM system, see Appendix \ref{appendix:AFM modes_thickness}.} 

\subsection{Hybrid modes and effect of the interaction with the antiferromagnet}

Our simulations also revealed that the higher order PSSW modes couple with the uniform FMR mode as the sample thickness increases, depicted in Figure \ref{thickness}(a). This phenomenon is observed in our simulations for the n=1 mode at 80-100 nm and between 140-160 nm in the case of the n=2 PSSW mode. Consequently, the mode profiles exhibit significant alterations with respect to the symmetrical n=1 and n=2 modes depicted for 50 nm (see Figure \ref{thickness}(c, d)), where there is no impact from the uniform mode.

\begin{figure}[t]
  \centering
  \includegraphics[width=1\linewidth]{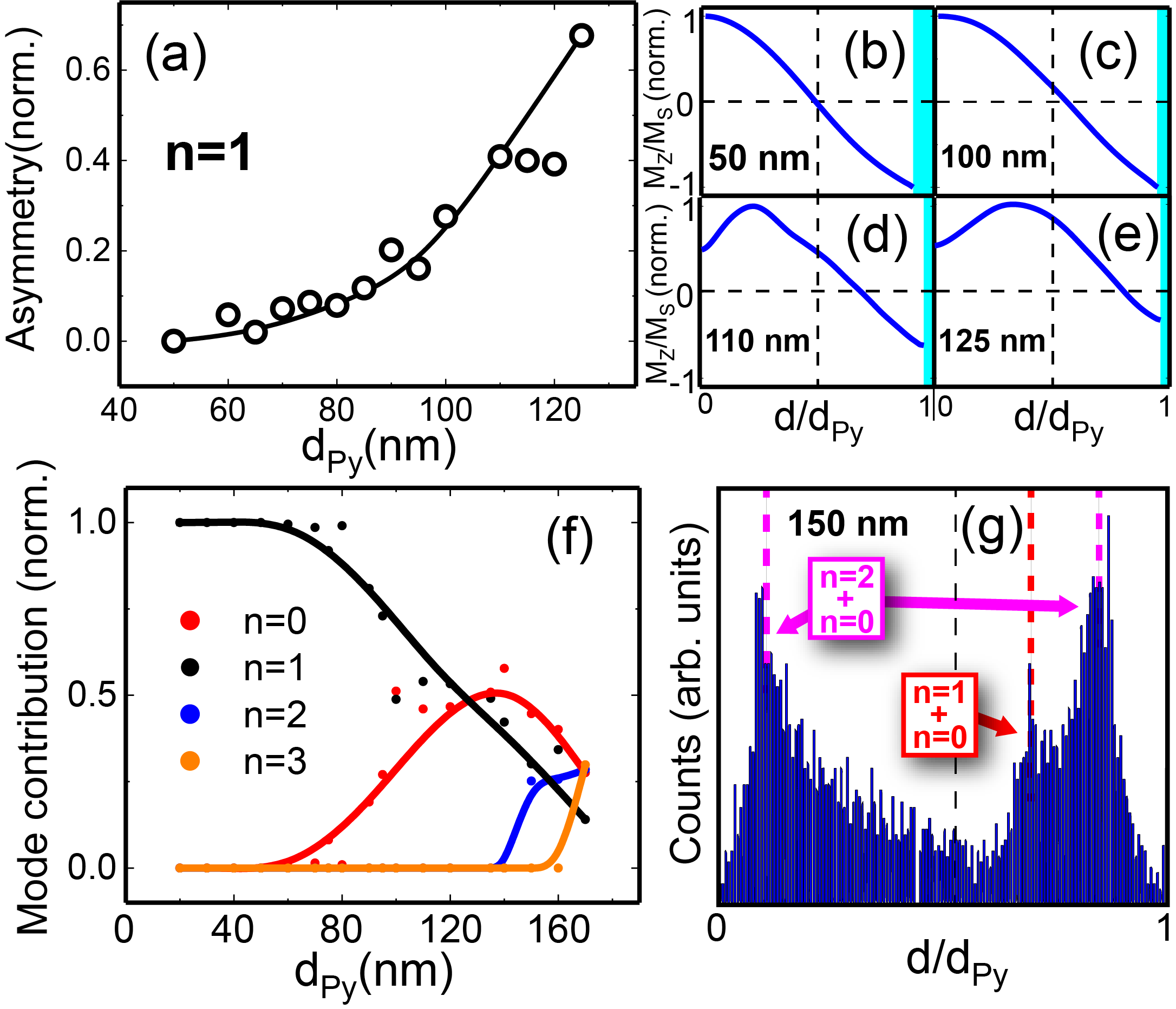} 
  \caption{(a) Asymmetry of the n=1 mode in relation with the thickness of the Py film. Asymmetry equal to 0 means a perfectly symmetrical placement of the spin wave node along the thickness of the film (perfect n=1). On the other hand, an asymmetry close to one means that the coupling of the fundamental and the n=1 mode have taken place, and that the mode profile is closer to that of the FMR mode. (b, c, d, e) are the one-dimensional mode profiles of the n=1 mode at 50, 100, 110 and 125 nm of thickness respectively. Clear blue parts at (b, c, d, e) correspond to the reduced M$_s$ layers in the film. (f) Mode contribution to the mode profiles obtained against the thickness of the film. (g) Histogram of the position of the node in a MW excitation of the main SW mode for a film thickness of 150 nm over 10 ns.}
  \label{asymmetry}
  \end{figure}

However, as the influence of the FMR mode becomes more pronounced, the mentioned asymmetry is more apparent. The nodes become displaced their symmetrical position, as is evident in Figure \ref{thickness}(f) for the n=1 in a 100 nm film and Figure \ref{thickness}(g) for the n=2 in a 150 nm sample. 
The FMR mode, seemingly reducing its frequency as the thickness increases, becomes similar to a n=1 mode profile for significant thicknesses, see Figure \ref{thickness}(e) for the mode profile on the 100 nm film. This mode profile is clearly distinct from the one observed when there is no coupling of the modes, shown on Figure \ref{thickness}(b).

We have investigated in more details the asymmetry (see Section \ref{methods} for explanation of the analysis) exhibited by the n=1 mode profile as the thickness of the film is increased. Figure \ref{asymmetry}(a) shows the asymmetry of the n=1 mode profile, i.e, a quantification of the shift from the symmetric position of the node. As the thickness of the simulated Py is increased, the asymmetry progressively increases, indicating a stronger influence of the FMR mode in relation to the n=1 mode. This evolution becomes evident in the cross-sections of the modes profiles depicted in Figures \ref{asymmetry}(b, c, d, e) for film thicknesses of 50, 100, 110 and 125 nm respectively. Beyond 50 nm, the n=1 mode gradually diverges from the symmetric position as the hybrid modes are generated, eventually resembling more and more the characteristic n=0 mode profile.

\begin{figure}[t]
  \centering
  \includegraphics[width=0.9\linewidth]{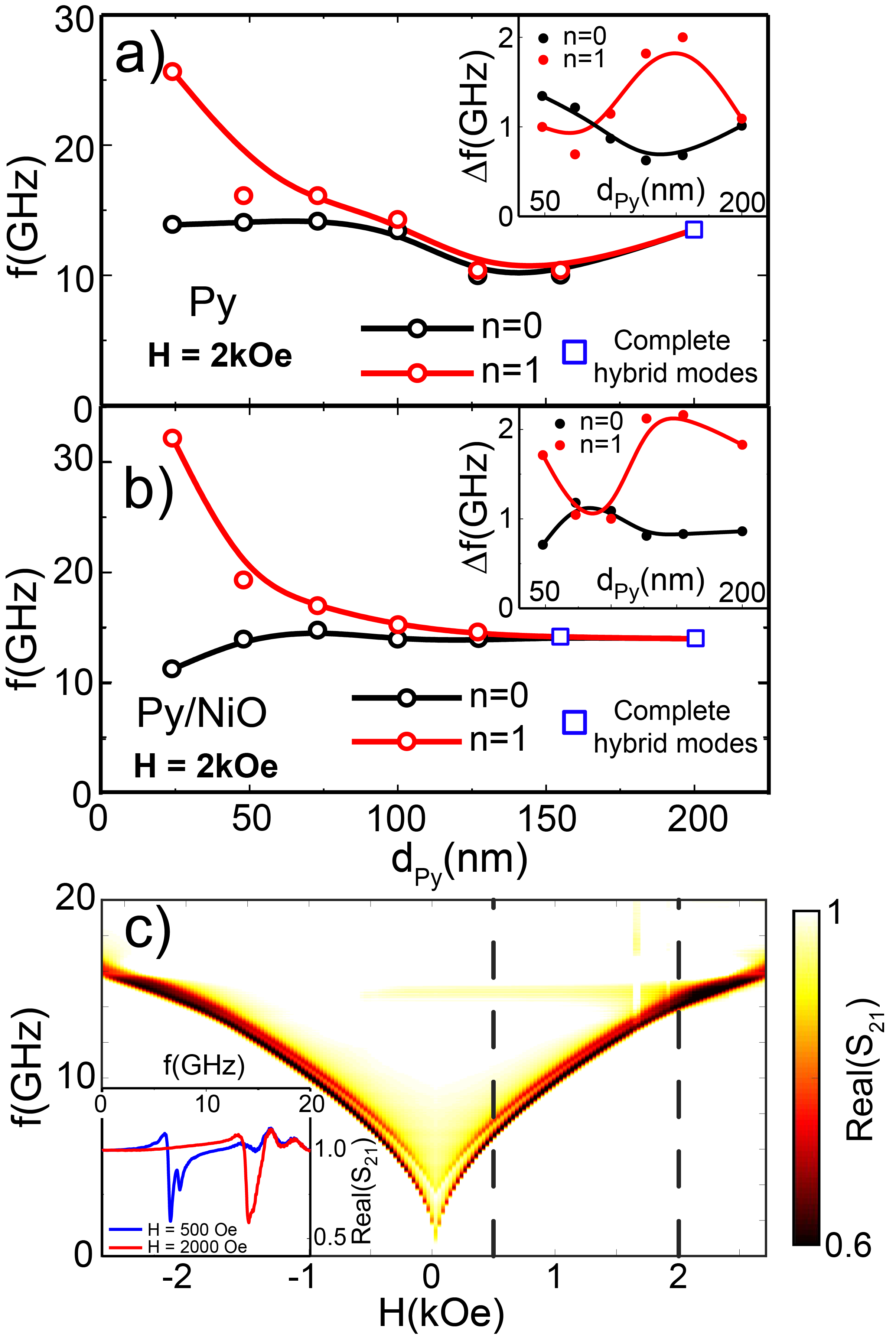} 
  \caption{(a) Resonance frequency of the n=0 uniform mode and n=1 against the thickness of control Py films for a bias field of 2kOe. (b) Same graph as panel (a), but for Py/NiO films. Insets in both panels represent the frequency FWHM of the modes. (c) Frequency-field map of the transmission in a Py(155nm)/NiO sample. Inset displays cross-sections at bias fields of 500 Oe and 2kOe.
  }
  \label{hybrid150nm}
  \end{figure}

The emergence of hybrid modes was evidenced not only through micromagnetic simulations, but also in VNA-FMR experiments at sufficiently high magnetic fields and for thicknesses above 120 nm (see Fig. \ref{hybrid150nm}(c)). Notably, the external in plane magnetic field helps the formation of hybrid modes; in the Py(155nm)/NiO sample the FMR uniform mode and the first order PSSW fully combine at fields above 1.8 kOe. Below 1 kOe, a distinct separation between them is observed. This is in line with the simulations performed, which demonstrate coupling of the modes at 2 kOe.

Experimentally, although the frequency shift is present upon the addition of the AFM, it is easier to hybridize the n=0 and n=1 modes in this case. This is reflected in Figures \ref{hybrid150nm}(a,b), where there is a complete combination of the two modes to form a hybrid mode in the case of Py(155 nm)/NiO. However, for just Py, the hybrid modes are not established in the same range of thicknesses.

\section{Conclusions}

In conclusion, we investigated experimentally and by simulations the static and dynamic magnetization in Py(\text{d$_{Py}$})/NiO(\diego{240} nm) and control Py(\text{d$_{Py}$}) films with thickness (\text{d$_{Py}$}) varied between 24 and 200 nm. We detected both perpendicular PSSWs spin waves and quasi uniform modes and found that in sufficiently thick Py films both exhibit a clear frequency shift when Py is exchange coupled to NiO. The frequency shift is attributed to the frozen magnetic moments at the Py/AFM interface, induced by the exchange bias, reducing the effective thickness of the ferromagnetic sample.

Additionally, we found that hybrid (n=0,1,2) PSSW modes emerge for sufficiently thick samples, which manifest (as detected by simulations) in asymmetric character modes with the n=0 mode influenced by the contribution of the different order PSSWs (n=1, 2...). The exchange coupling between Py and NiO apparently stimulates such hybrid modes to emerge at reduced (in comparison with control Py films) thickness. \diegonew{Our findings are expected to provide a deeper understanding of the relation between exchange coupling and magnetization dynamics in ferromagnet-antiferromagnet hybrids. Overall, our results will be important for the applications of AFM/FM heterostructures which hold great promise for enabling ultrafast spintronics devices with low power consumption, high efficiency, as well as novel functionalities, paving the way for the next generation of information processing and storage technologies.}

\section{acknowledgments}
The work in Madrid was supported by Spanish Ministry of Science and Innovation (PID2021-124585NB-C32, TED2021-130196B-C22) and Consejería de Educación e Investigación de la Comunidad de Madrid (NANOMAGCOST-CM Ref. P2018/NMT-4321) grants. F.G.A. acknowledges financial support from the Spanish Ministry of Science and Innovation, through the María de Maeztu Programme for Units of Excellence in R\&D (CEX2018-000805-M) and "Acción financiada por la Comunidad de Madrid en el marco del convenio plurianual con la Universidad Autónoma de Madrid en Linea 3: Excelencia para el Profesorado Universitario". \diego{We acknowledge the Spanish Ministerio de Ciencia, Innovación y Universidades and the Consejo Superior de Investigaciones Científicas for provision of synchrotron radiation at the BM25-SpLine beamline at The European Synchrotron.}

\clearpage
\section*{Appendix}
\appendix

\section{Sample preparation and structural characterization}
\label{appendix: characterization}

NiO(AFM)/Fe$_{20}$Ni$_{80}$(FM) bilayers were deposited on Al$_2$O$_3$ (0001) substrates by ion beam sputtering using Ar$^+$ ions from a 3 cm Kaufmann-type ion source in a vacuum chamber with a base pressure of 2$\times$10$^{-5}$ Pa. The thin nickel oxide films were obtained from a pure nickel (99.99$\%$) target in a controlled atmosphere of oxygen and argon with P(O$_2$) = 2.6$\times$10$^{-2}$ Pa and P(Ar)= 1.4$\times$10$^{-2}$ Pa. The substrate temperature during NiO deposition was maintained at 500$^{\circ}$C and the sputter ion energy and current density were 650 eV and 1.1 mA/cm$^2$. The thickness of the NiO layer was high enough (240$\pm$20 nm) to avoid AFM thickness effect. Fe$_{20}$Ni$_{80}$ thin films with thicknesses ranging between 4 and \diego{200 nm} have been obtained from a 99.99$\%$ pure Fe$_{20}$Ni$_{80}$ target in an Ar atmosphere with P(Ar)= 1.4$\times$10$^{-2}$ Pa and 100$^{\circ}$C substrate temperature. In this case, the sputter ion energy and current density were 550 eV and 0.8 mA/cm$^2$ respectively. The substrates rotated at 2 rpm to increase the homogeneity of the deposit. An Al capping layer 2 nm thick was deposited on top of the FM layer to prevent oxidation for “ex situ” studies. 
The crystal structure and texture of the different films was analyzed by XRD (X-ray difraction) in a $\theta$/2$\theta$ configuration using a PANanalytical X'Pert MPD system with Cu-K$\alpha$ radiation. \diego{XRR (X-ray reflectivity) and GIXRD (grazing angle XRD) were also performed at ESRF Synchrotron (European Synchrotron Radiation Facility) at the BM25-Spline beamline. In-plane and out-of-plane measurements were performed at room temperature and ambient pressure conditions using a six-circle diffractometer in vertical geometry to profit from the high collimation of the synchrotron beam. A photon energy of 15keV (0.0827 nm) was used.}
In-depth composition and thickness of the different layers were determined by RBS at the CMAM 5 MV tandem accelerator using 4He$^+$ at 1.8 MeV \cite{RedondoCubero2021} A silicon barrier detector, at a scattering angle of 170.5$^{\circ}$, measured the backscattering ions while a three-axis goniometer was employed to control the crystal position. The elemental distribution and thickness quantification of the different layers were determined with the SIMNRA simulation software package.

\begin{figure}[t]
  \centering
  \includegraphics[width=0.85\linewidth]{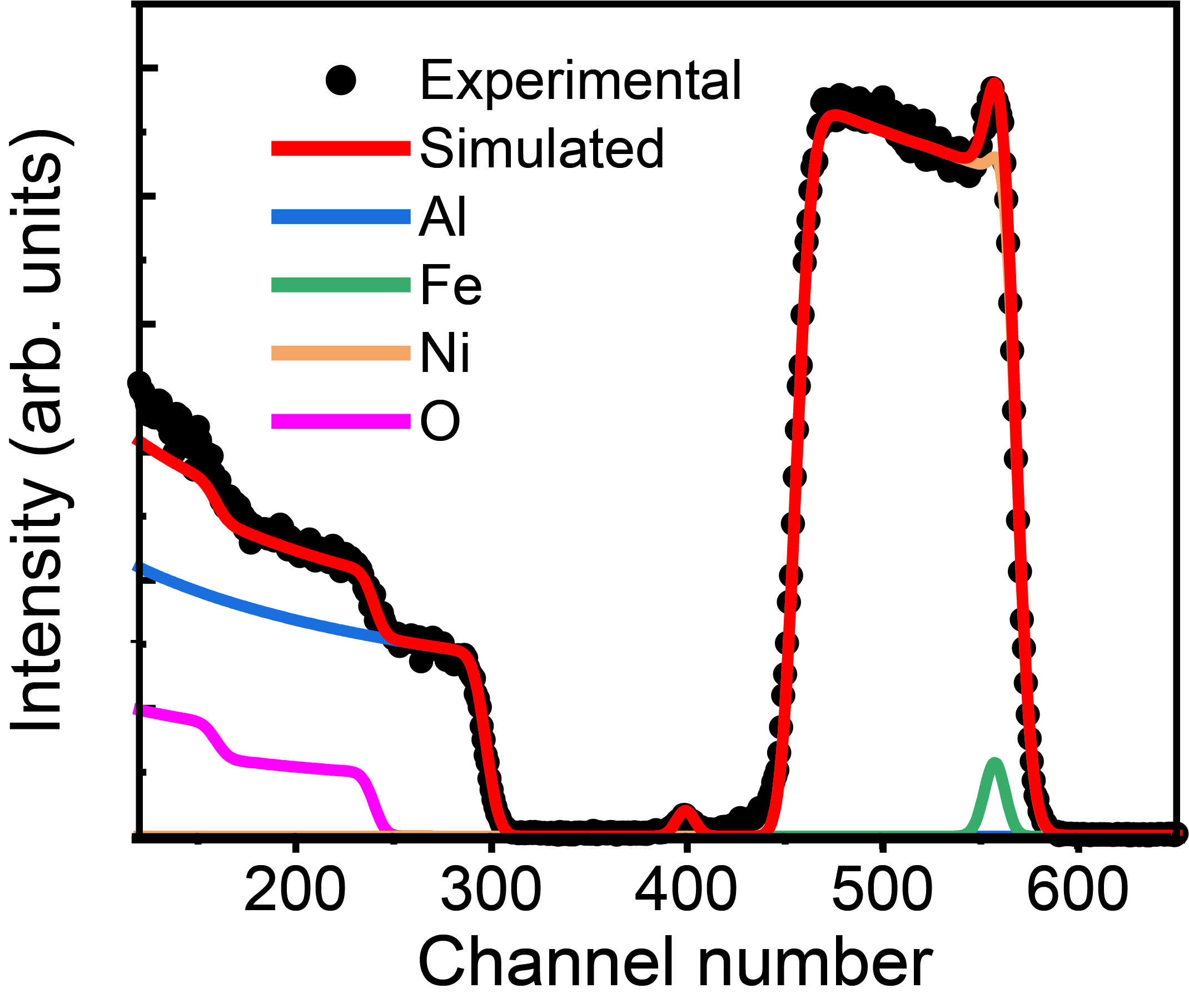} 
  \caption{Random RBS spectra of a NiO/Py bilayer obtained with 4He$^+$ ions at 1.8 MeV and its simulation. The fitting signal for O, Ni, Al and Fe are included.}
  \label{sample1}
  \end{figure}

\begin{figure}[t]
  \centering
  \includegraphics[width=1\linewidth]{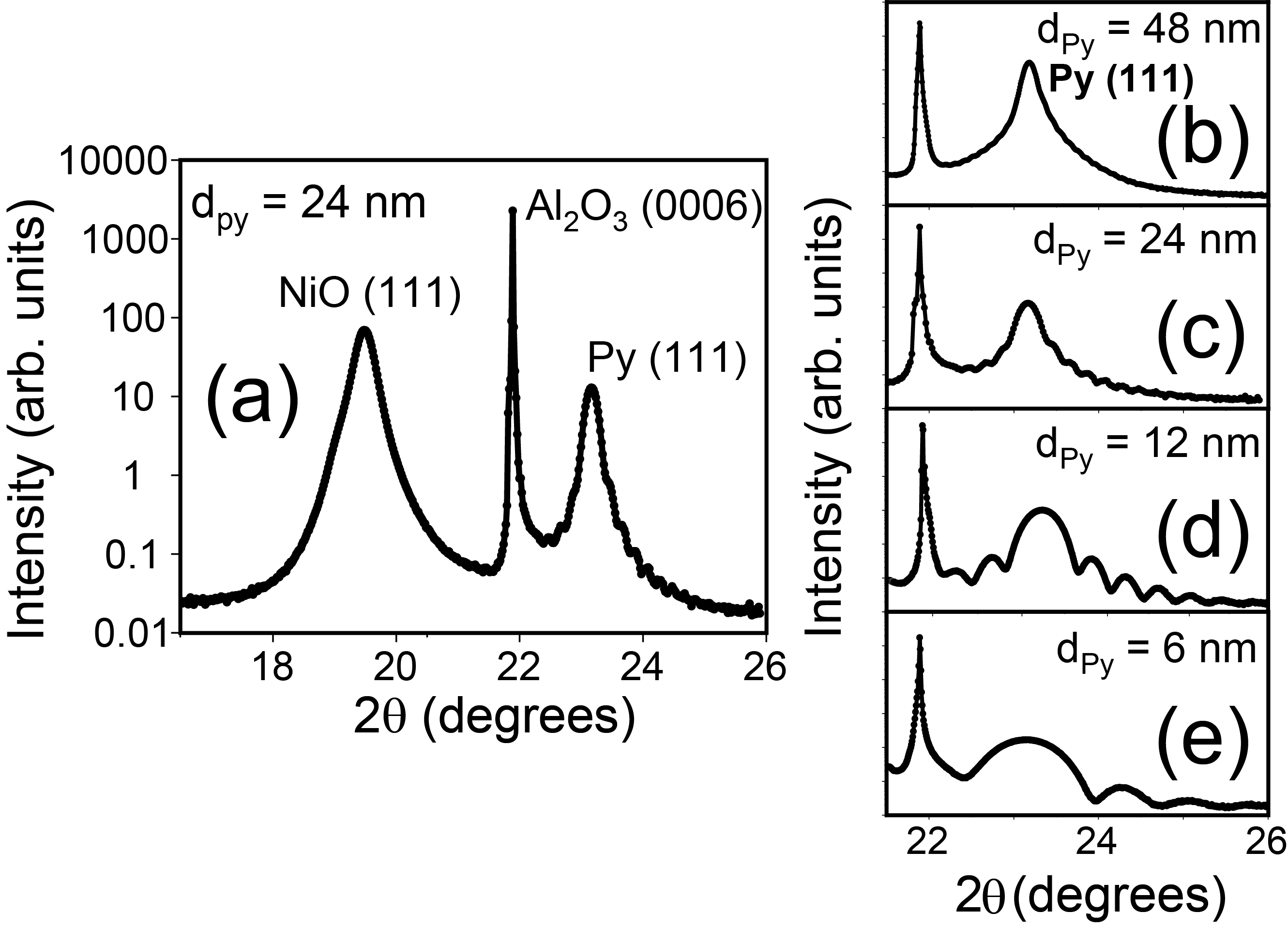} 
  \caption{(a) complete XRD $\theta$/2$\theta$  scan at a photon energy of 15 KeV of a NiO/Py bilayer (d$_{Py}$ = 12 nm) (b, c, d, e) XRD $\theta$/2$\theta$  scans of the (0006) Al$_2$O$_3$ and (111) Py diffraction peaks for d$_{Py}$= 48, 24, 12 and 6 nm respectively.}
  \label{sample2}
  \end{figure}
  
The angular dependence of magnetization was investigated at room temperature by high-resolution vectorial-Kerr magneto-optical measurements (MOKE) in longitudinal configuration. MOKE hysteresis loops were recorded by changing the in-plane angular orientation of the sample, keeping fixed the external magnetic field direction. The angular orientation, ranging from 0 to 360$^{\circ}$ was probed at intervals of 9$^{\circ}$ with a maximum applied magnetic field of 1.1 kOe.

\begin{figure}[t]
  \centering
  \includegraphics[width=0.75\linewidth]{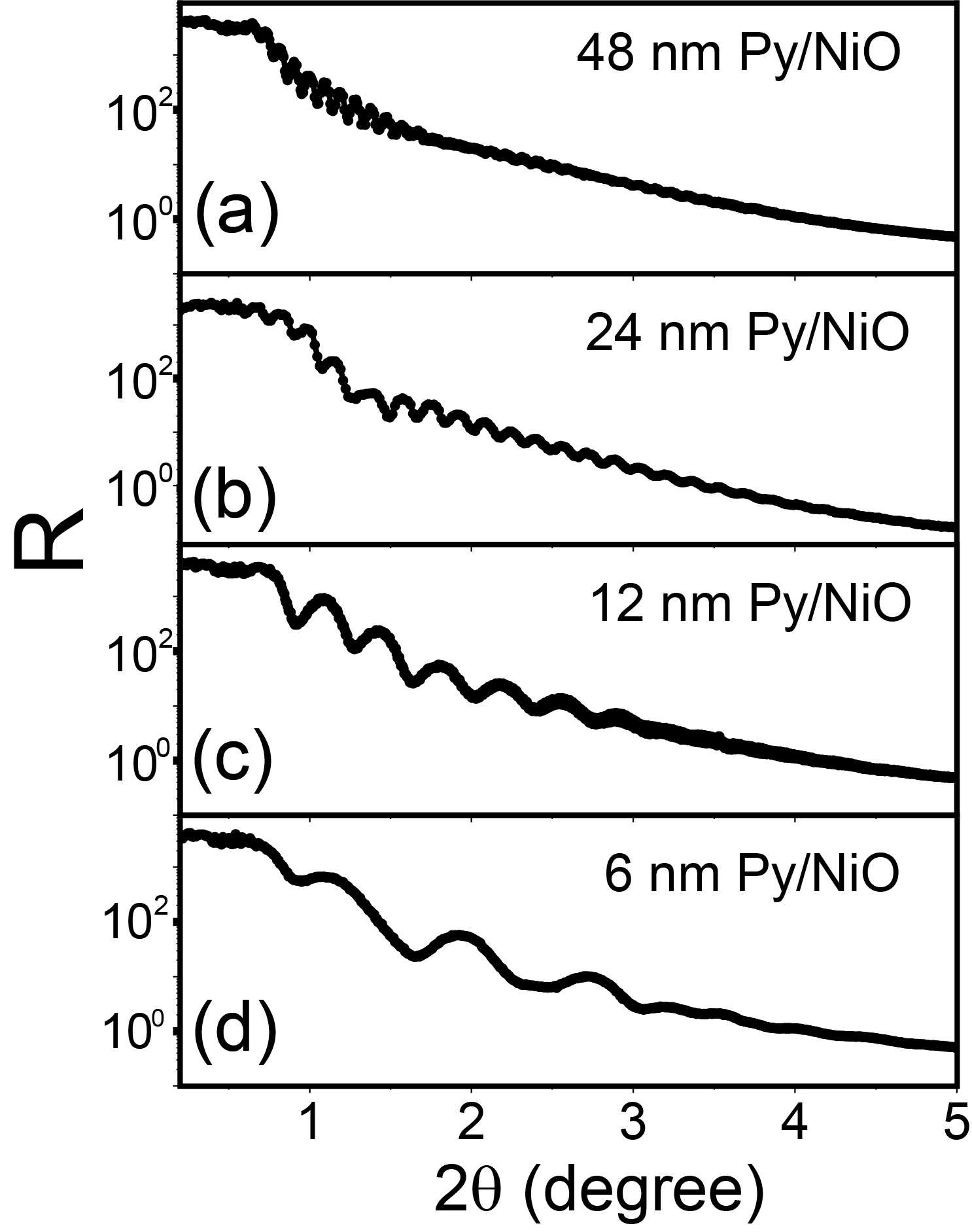} 
  \caption{Low angle reflectivity for different NiO/Py bilayers grown on Al$_2$O$_3$ (0001) substrates.}
  \label{sample3}
  \end{figure}

\begin{figure}[t]
  \centering
  \includegraphics[width=1\linewidth]{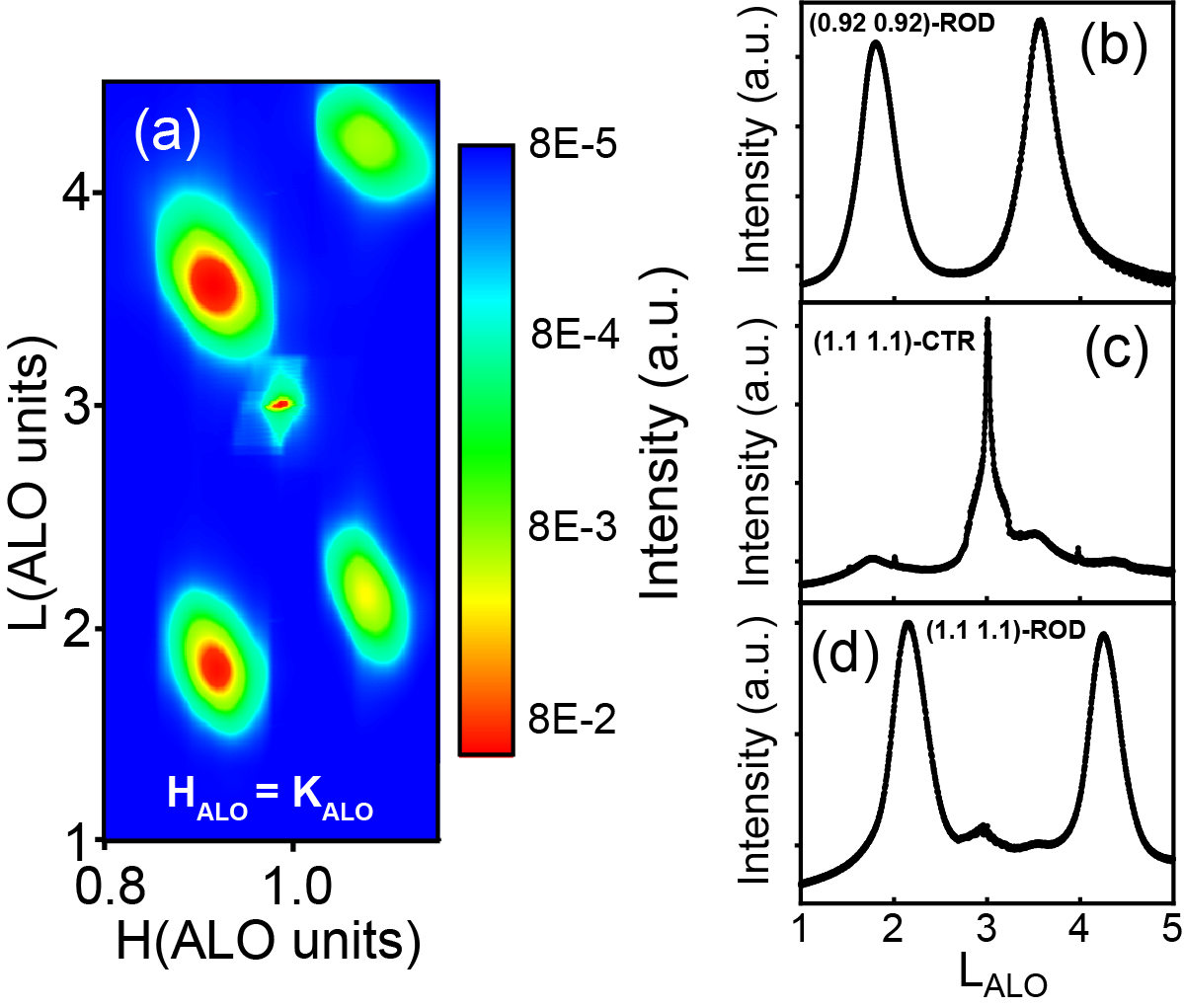} 
  \caption{(a) H = K reciprocal space maps for NiO/Py bilayer (d$_{Py}$ = 24 nm) grown Al$_2$O$_3$ (0001) substrate. (b, c, d) CTR and RODs scans obtained for H = K = 0.92, H = K = 1 and H = K = 1.1.}
  \label{sample4}
  \end{figure}

In-depth elemental composition and thickness of the different Py layers as well as the complete heterostructure NiO/Py/Al capping were determined by Rutherford backscattering spectroscopy (RBS) in a 5 MV tandem accelerator using 4He$^+$ at 1.8 MeV.  As an example, the experimental spectrum of a NiO/Py bilayer have been shown on Figure \ref{sample1} that also include the fitting by using SIMNRA simulation program assuming a three layer model, i.e., NiO, Py and Al layers. The estimated composition and thickness for this particular sample is Ni$_{0.49}$O$_{0.51}$ (d = 223 nm)/Fe$_{0.19}$O$_{0.81}$ (d = 11.6 nm)/Al (d = 2 nm). 
Figure \ref{sample2}(a) shows the $\theta$/2$\theta$ XRD scans of a NiO/Py bilayer grown on (0001) Al$_2$O$_3$ substrates obtained with a photon energy of 15 KeV (0.0827 nm). The diffraction pattern shows the NiO (111) diffraction peak at 19.49$^{\circ}$, the only one from NiO that can be distinguished in the complete diffractogram, the Al$_2$O$_3$ (0006) peak at 21.89$^{\circ}$ and the Py (111) peak at 23.17$^{\circ}$ \diego{that indicate a single oriented growth along the 111 direction}. Fig. \ref{sample2}(b) shows just the Al$_2$O$_3$ (0006) peak and the Py (111) peak for bilayers with different Py thickness. It can be observed in the kissing fringes around the Py (111) film Bragg peak, indicating a high quality with smooth surfaces and abrupt interfaces. With the help of the position of successive Kiessig fringes, it is possible to determine the Py film thickness, with results in good agreement with the obtained by RBS quantification.
Figure \ref{sample3} displays the low-angle XRR result for different NiO/Py bilayers deposited on Al$_2$O$_3$ (0001) substrates. The presence of intense Kiessig fringes, which result from interference of the X-ray beams reflected on the film surface and on the interface to the NiO layer, is also observed in the reflectivity data, indicating a sharp NiO/Py interface and smooth Py surface. \diego{The oscillations observed allow also an estimation of the thickness of the ferromagnetic Py layer in good agreement with the ones obtained by RBS. In Fig. \ref{sample3}(c), it is also possible to observe very small oscillations that allow to estimate the thickness of the NiO layer for this particular sample with a value of 230 nm.}
\diego{Fig. \ref{sample4} shows representative reciprocal space map for a NiO/Py bilayer with d$_{Py}$ = 24 nm obtained by off-specular GIXRD using synchrotron beam. An incommensurate epitaxial growth is evidenced from the non-coincidence of the Py and NiO layer in-plane diffraction peaks respect to those of the Al$_2$O$_3$ (0001) substrate. The substrate in-plane diffraction peak is obtained at an integer in-plane H/K value, as the orientation matrix has been defined using the substrate (ALO) lattice parameters, while the NiO and Py in-plane diffraction peaks are obtained at H=K=0.92 and 1.1 substrate reciprocal lattice units, respectively. This XRD pattern can be explained by a non–coincidence epitaxial growth in which the in-plane lattice of the single oriented NiO/Py (111) rotates 30$^{\circ}$ respect to the underlying Al$_2$O$_3$ (0001) plane in order to reduce the lattice mismatch minimizing the epitaxial energy. Calculations reveal an in–plane lattice parameter of 4.22(4) \r{A} (cubic notation) for the NiO and 3.55(1) \r{A} (cubic notation) for the Py. These results indicate that the NiO grows with a tensile stress while the Py grows fully relaxed. The line profile scans along the in-plane substrate H/K value (CTR) and the corresponding to the NiO and Py (ROD) are also shown. A c-lattice parameter of 4.20(2) and 3.53(2) A are obtained for NiO and Py, respectively.}

\begin{figure}[t]
  \centering
  \includegraphics[width=0.9\linewidth]{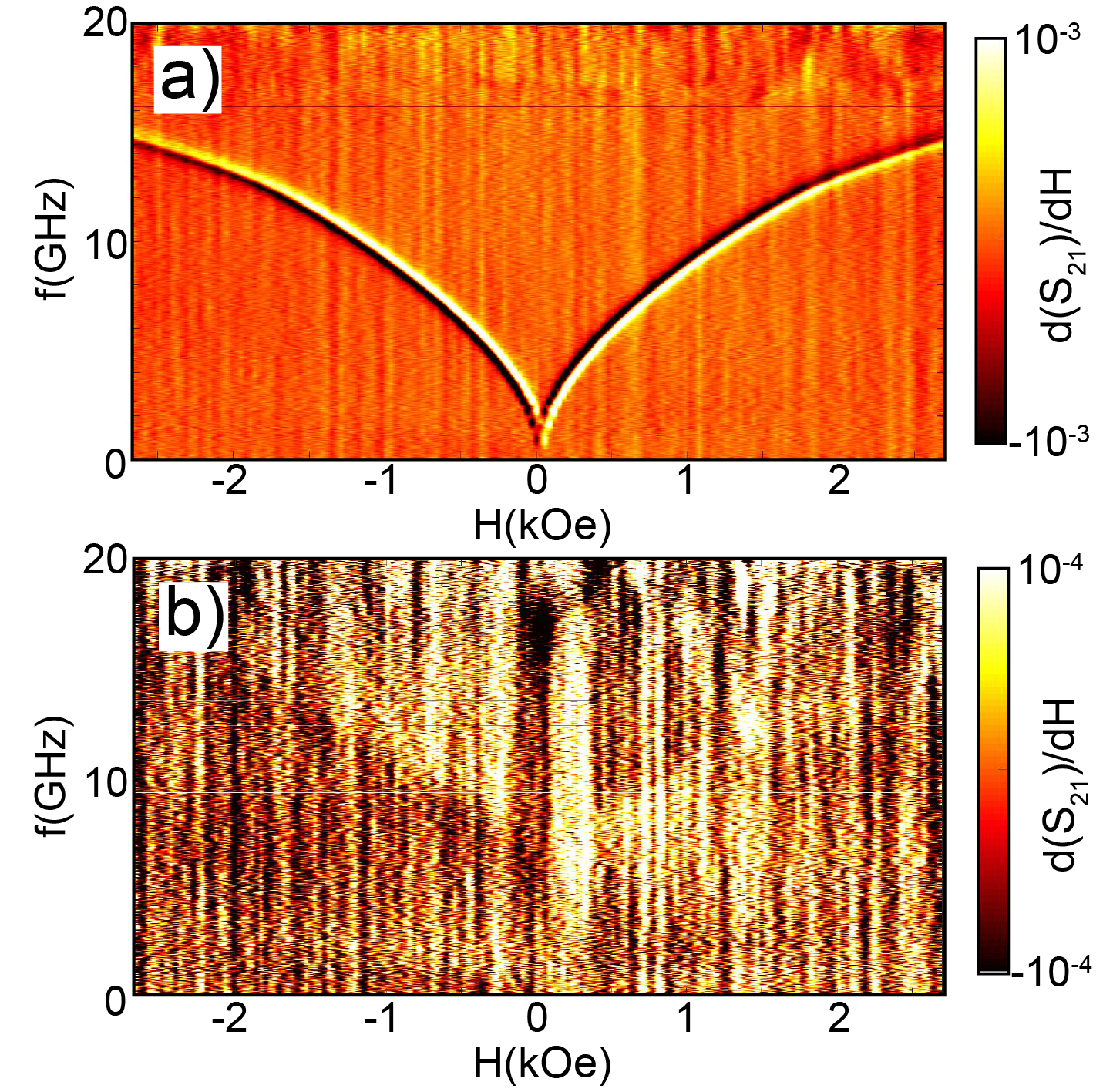} 
  \caption{(a) Py(4nm) and (b) Py(4nm)/NiO VNA-FMR results of the field derivative of the transmission implemented as frequency-field maps. Note that the colormap scale range in (b) has been deliberately decreased to appreciate the low FMR signal.}
  \label{low_thickness}
  \end{figure}

\section{Low thickness VNA-FMR measurements}

\label{appendix:low_thickness}

We conducted VNA-FMR experiments to analyze the dynamic response of extremely thin Py and Py/NiO films. In Figure \ref{low_thickness}(a), a 4 nm thick Py measurement is presented, where the uniform mode is clearly observed. However, when the same nominal thickness Py film is coupled with NiO, the signal coming from the FMR mode signal greatly diminished (Figure \ref{low_thickness}(b)). 
 
This result provides further validation of the effective film thickness reduction when coupling an AFM and an F, a consequence of the exchange bias freezing the magnetic moments of the Py at the interface. This, in turn, leads to an enhancement of the frequency of the PSSW modes. Our calculations indicate that the frozen moments in our films reduce the effective thickness by just over 2 nm upon the NiO addition, a value that remains largely consistent across all of our films (see Figure \ref{Shift}(b)). Hence, these dynamic measurements align with the expected outcomes for films as thin as these, where a significant portion of the thickness is reduced, and the signal of the ferromagnetic modes is nearly completely suppressed.



\section{Resolved standing spin waves modes above n=2}

\label{appendix:nabove2}
\begin{figure}[t]
  \centering
  \includegraphics[width=1\linewidth]{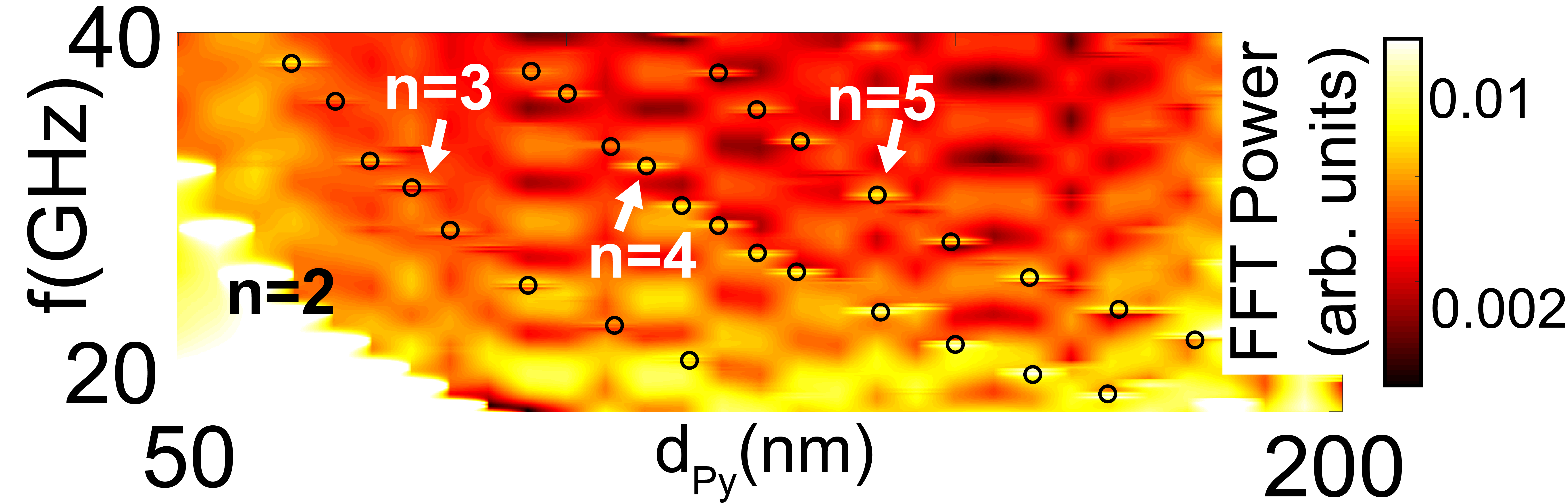} 
  \caption{Evolution of the PSSW modes above n=2 at a constant in-plane magnetic field of H = 2 kOe.}
  \label{n2}
  \end{figure}

The resolved main SW modes on our micromagnetic simulations outcome clearly showed strong PSSWs dynamics corresponding to n=0, 1 and 2  (see Figure \ref{thickness} in the main text). However, an in-depth examination of the dynamics of the system indicated weak modes above the second order PSSW as the thickness of the film increases, as shown in Figure \ref{n2}. These modes, distinguishable up until n=5, also exhibit a similar pattern as the lower order PSSWs (f(GHz)= A + B/d$_{Py}^2$(nm)).  

\section{Simulations on the exchange bias along the thickness of the film}
\label{appendix:exch bias}

\begin{figure}[t]
  \centering
  \includegraphics[width=0.8\linewidth]{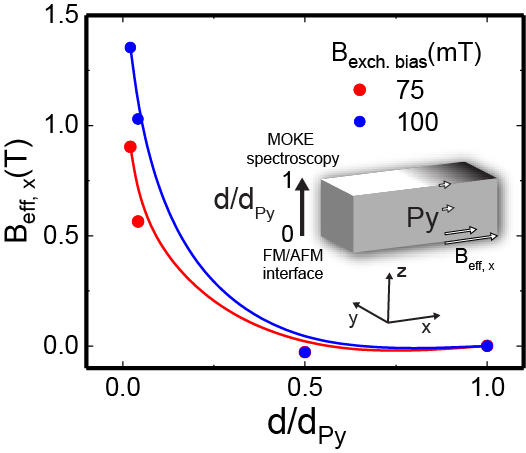} 
  \caption{Thickness evolution of the effective field in the exchange bias direction on a 48 nm Py simulated sample. The exchange bias is applied at the interface (d/d$_{Py}$=0).
  }
  \label{exch. bias}
  \end{figure}

As mentioned in section \ref{optimization}, the difference between the values obtained of the exchange bias via MOKE (see Figure \ref{hysteresis}(c)), on the tenths or hundreds of Oe, and the applied exchange bias at the interface, 1 kOe, are distant in value. As Figure \ref{exch. bias} shows via simulations, this order of magnitude is recovered due to the different location of the field detected/applied: MOKE spectroscopy on top of the Py film, whereas the FM/AFM coupling that provides the exchange bias, on the interface. 

These simulations are done on a Py(48nm) sample and the exchange bias field is applied on the first 2 nm of the sample, which is coherent with the reduced effective thickness calculations displayed on Figure \ref{Shift}(b).

\section{Evolution of the SW modes in a Py/AFM bilayer with the thickness of the film}
\label{appendix:AFM modes_thickness}

\begin{figure}[t]
  \centering
  \includegraphics[width=1\linewidth]{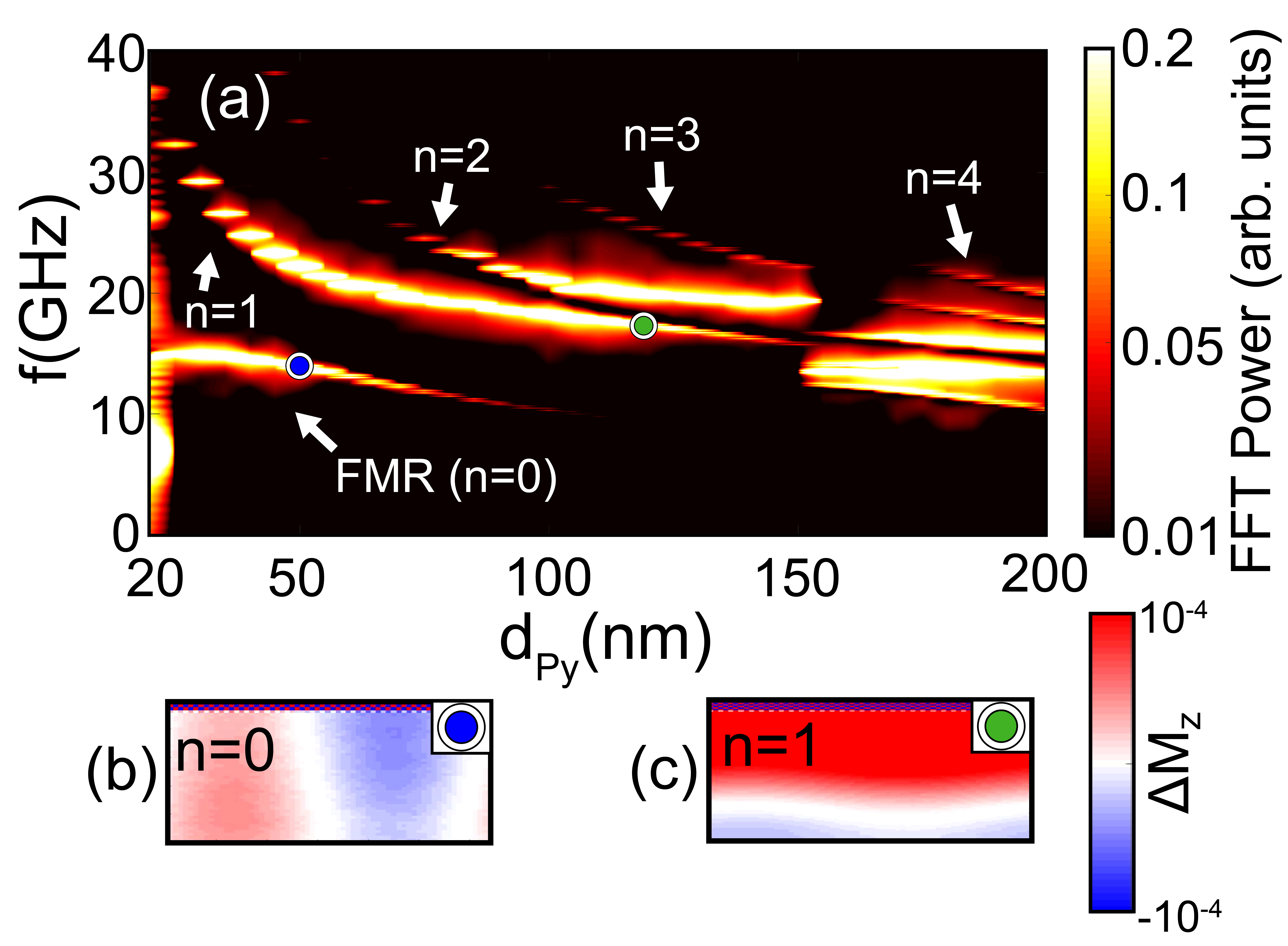} 
  \caption{(a) Evolution of the primary SW mode frequencies at a constant magnetic field of H = 2 kOe as the Py layer thickness varies from 20 nm to 200 nm in a Py/AFM bilayer. The FFT power is normalized at each film thickness.}
  \label{thicknessNiO}
  \end{figure}
Adding an AFM/FM interface with K$_u$ = 10$^{-5}$ J/m$^3$ and an exchange bias field of H$_{eb}$ = 1 kOe, as suggested by the parameter optimization described in Section \ref{optimization} of the main text, leads to noticeable shifts in the frequencies of the PSSWs modes for low thicknesses (refer to Figure \ref{thicknessNiO}) in comparison with the Py system lacking the AFM/FM interface depicted in Figure \ref{thickness}(a). These frequency shifts are towards higher frequencies, consistent with our experimental findings using the VNA-FMR system collected in Figure \ref{PSSW}(a). 

The resolved antisymmetric modes, particularly n=3, exhibit significantly greater amplitude compared to the Py system. This amplification is attributed to the induced out-of-plane asymmetry caused by the uniaxial anisotropy and the exchange bias of the system, which are localized on the bottom layer where the AFM is placed. Consequently, the applied sinc-shaped magnetic pulse excites the antisymetric modes more effectively. Another interesting characteristic of the resolved modes is that at larger thicknesses of the Py layer (150-165 nm), there is an abrupt shift towards lower frequencies, suggesting a diminished impact of the AFM. In these cases, the influence of the uniaxial anisotropy and the exchange bias is reduced. The precise values of these interface parameters determine the sharp, step-like transition observed in the modes.

\bibliography{bibliography}

\end{document}